\documentclass[preprint,preprintnumbers,amsmath,amssymb, superscriptaddress,letterpaper, pre]{revtex4}
\usepackage{graphicx}
\usepackage{subfigure}
\usepackage{amsmath}
\usepackage{amssymb}
\usepackage{dcolumn}
\usepackage{bm}

\begin{document}

\title{Phase Behavior of Colloidal Superballs: Shape Interpolation from Spheres to Cubes}
\author{Robert D. Batten}
\affiliation{Department of Chemical Engineering, Princeton 
University, Princeton, NJ 08544, USA}

\author{Frank H. Stillinger} 
\affiliation{Department of Chemistry,
Princeton University, Princeton, NJ, 08544 USA}

\author{Salvatore Torquato} 
\affiliation{Department of Chemistry, Princeton University, Princeton, NJ, 08544
USA} 
\affiliation{Department of Physics, Princeton University, Princeton, NJ, 08544
USA}
\affiliation{Princeton Center for Theoretical Science, Princeton University,
Princeton, NJ 08544, USA}
\affiliation{Princeton Institute for the Science and Technology of Materials, Princeton University, Princeton, NJ
08544, USA}
\affiliation{Program in Applied and Computational Mathematics, Princeton
University, Princeton, NJ 08544, USA} 
\affiliation{School of Natural Sciences, Institute for Advanced Study,
Princeton, NJ, 08544 USA}

\altaffiliation{Corresponding author}
\email{torquato@electron.princeton.edu} \date{\today}

\date{\today}

\begin{abstract}
The phase behavior of hard superballs is examined using molecular dynamics
within a deformable periodic simulation box. A superball's interior is defined
by the inequality $|x|^{2q} + |y|^{2q} + |z|^{2q} \leq 1$, which provides a
versatile family of convex particles ($q\geq0.5$) with cube-like and
octahedron-like shapes as well as concave particles ($q<0.5$) with
octahedron-like shapes.  Here, we consider the convex case with a deformation
parameter $q$ between the sphere point ($q=1$) and the cube ($q=\infty$). We
find that the asphericity plays a significant role in the extent of cubatic
ordering of both the liquid and crystal phases.  Calculation of the first few
virial coefficients shows that superballs that are visually similar to cubes can
have low-density equations of state closer to spheres than to cubes.  Dense
liquids of superballs display cubatic orientational order that extends over
several particle lengths only for large $q$.  Along the ordered, high-density
equation of state, superballs with $1 < q < 3$ exhibit clear evidence of a phase
transition from a crystal state to a state with reduced long-ranged
orientational order upon the reduction of density. For $q\geq 3$, long-ranged
orientational order persists until the melting transition. The width of
coexistence region between the liquid and ordered, high-density phase decreases
with $q$ up to $q=4.0$.  The structures of the high-density phases are examined
using certain order parameters, distribution functions, and orientational
correlation functions. We also find that a fixed simulation cell induces
artificial phase transitions that are out of equilibrium.  Current fabrication
techniques allow for the synthesis of colloidal superballs, and thus the phase
behavior of such systems can be investigated experimentally. 

\end{abstract}

\maketitle

\section{Introduction}

As the ability to control size, shape, and structure of nanoparticles
\cite{sun2002shape, grzelczak2008shape, tao2008shape} and colloids
\cite{brown2000fabricating, zhao2007nematic, snoeks2000colloidal} improves,
computer simulation and theory of hard-particle systems becomes increasingly
important to the identification of technologically useful properties. Hard
convex particles have been used as models for simple atomic liquids and solids
as a means to connect the particle shape and excluded volume to the entropy and
to the equilibrium phase diagram of a system. The hard-sphere model has a rich
history and continues to provide deep insights into fundamental physical
phenomena \cite{kirkwood1950radial,metropolis1953equation, alder1957phase,
ree1964fifth, hoover1968mtc, carnahan1969equation, speedy1998pressure,
torquato2002rhm, skoge2006packing}. However, nonspherical hard particles exhibit
more complex phase behavior than hard spheres, since the possibility of
anisotropic phases arises, including smectic, nematic, columnar, and cubatic
liquid crystals \cite{allen1993hcb}.

The cubatic phase has garnered recent attention because, unlike other
liquid-crystalline phases, it is characterized by ordering in three mutually
perpendicular directions while the particles retain translational mobility
\cite{john2004clc}. Such unusual ordering may lead to novel optical,
rheological, or transport properties. The cubatic phase has been discovered in
several hard-particle systems.  Cut hard spheres of certain aspect ratios form
small stacks that align perpendicularly to neighboring stacks
\cite{veerman1992phase}.  The Onsager cross, a particle consisting of three thin
rods aligned along orthogonal axes and intersecting at their midpoints
\cite{blaak1998phase} and tetrapods, hard bodies formed by four rods connected
at tetrahedral angles \cite{blaak2004cubatic}, are examples of nonconvex
particles that exhibit cubatic ordering.  The cubatic phase also arises in
systems of perfect tetragonal parallelepipeds \cite{john2008pbc} as well systems
of cuboids, nonconvex particles consisting of an array of tangent hard spheres
that approximate tetragonal parallelepipeds  \cite{john2004clc, john2005phase}.
Perfect tetragonal parallelepipeds have sharp corners and flat faces while the
cuboid is ``bumpy'' to approximate friction. Monte Carlo simulation studies of
these particles revealed a cubatic phase, or parquet phase for aspects ratio
other than 1:1:1, that arises between the liquid and crystal phases
\cite{john2004clc, john2005phase, john2008pbc}.

\begin{figure} 
\subfigure[ $\text{ }q = 1.0$]{\label{fig:p1.0.triad}
\includegraphics[width=0.18\textwidth]{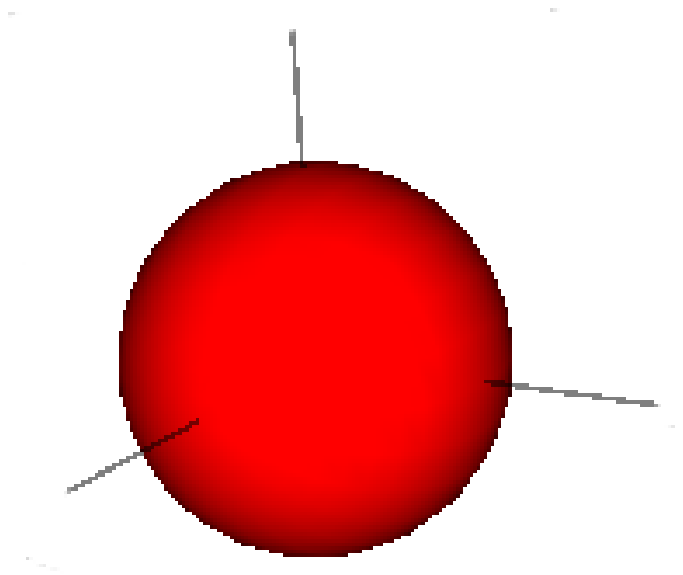}} 
\subfigure[ $\text{ }q = 1.5$]{\label{fig:p1.5.triad}
\includegraphics[width=0.18\textwidth]{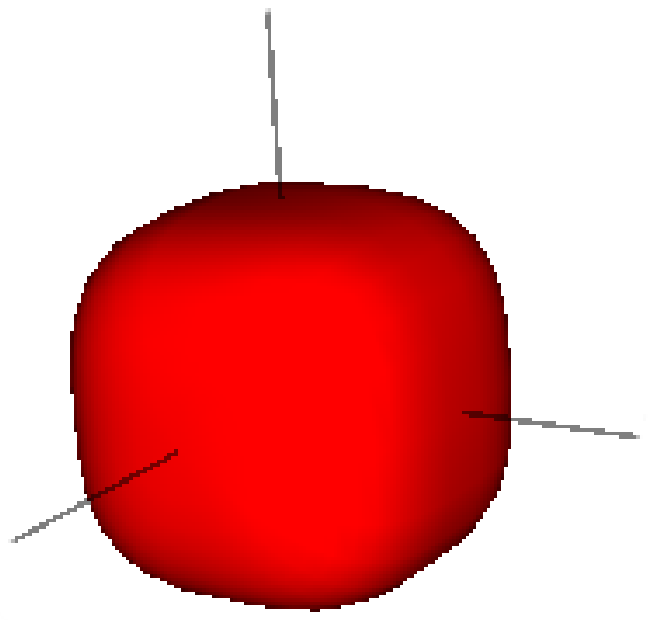}} 
\subfigure[ $\text{ }q = 2.0$]{\label{fig:p2.0.triad}
\includegraphics[width=0.18\textwidth]{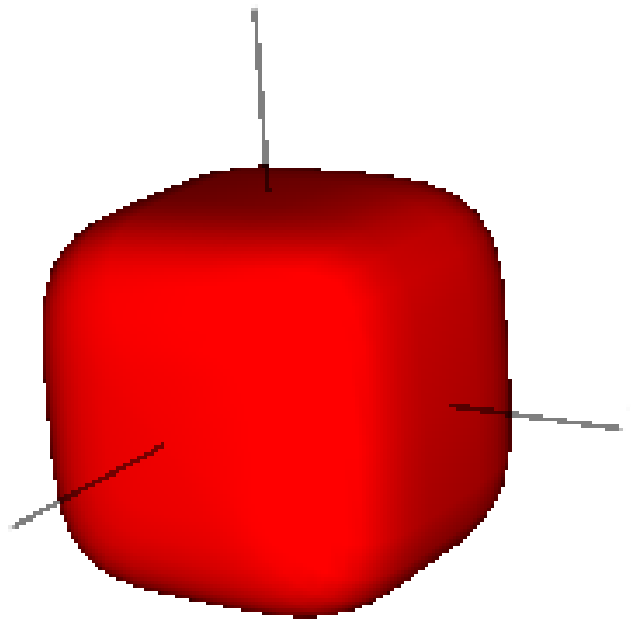}} 
\subfigure[ $\text{ }q = 4.0$]{\label{fig:p4.0.triad}
\includegraphics[width=0.18\textwidth]{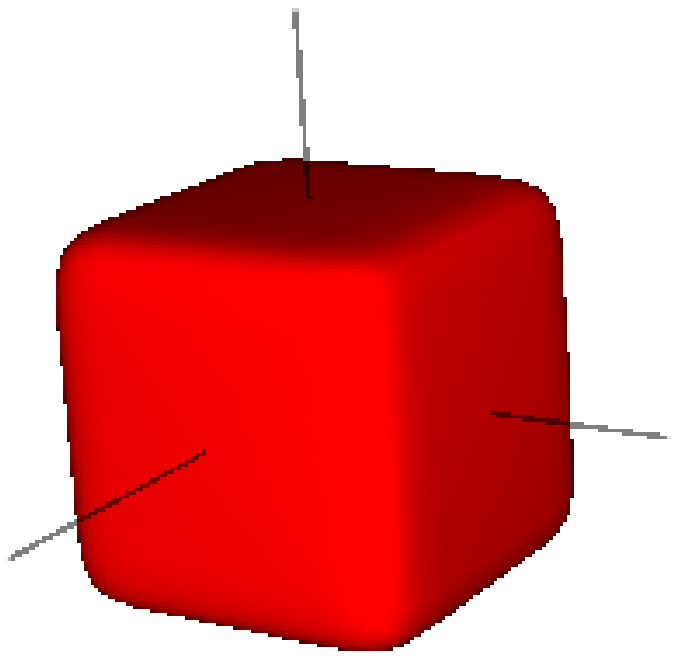}} 
\subfigure[ $\text{ }q = \infty$]{\label{fig:pinf.triad}
\includegraphics[width=0.18\textwidth]{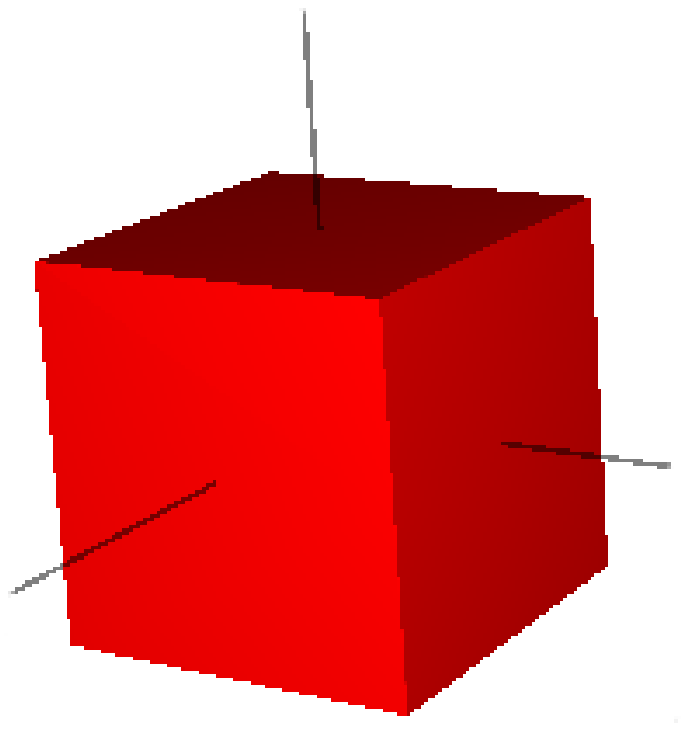}}
\caption{(Color online) Superballs for certain deformation parameter $q$.  The
lines represent the three equivalent principal axes.}
\label{fig:superballs}
\end{figure}

In this paper, we use molecular dynamics (MD) to investigate the equilibrium
phase behavior and the onset of cubatic ordering in systems of cube-like
superballs. A superball is a centrally symmetric particle defined by
\cite{jiao2009ops} 
\begin{equation} 
|x|^{2q} + |y|^{2q} + |z|^{2q} \leq 1, 
\end{equation} 
where $x,y$ and $z$ are Cartesian coordinates and $q$ is the deformation
parameter \cite{footnote2}.  Superballs can take on concave shapes ranging from
a cross ($q=0$) to the convex octahedron ($q=1/2$) to a sphere ($q=1$) and
finally to a cube ($q=\infty$). We focus on the ``cube-like'' range $1 \leq q
\leq \infty$ to examine the interpolation from spheres to cubes and reveal the
role of particle shape and curvature on cubatic ordering. As $q$ is increased
from unity, the particle takes on more cube-like characteristics as edges and
corners sharpen while faces flatten. Figure \ref{fig:superballs} displays
several superballs with their principal axes for the deformation from unity to
infinity. 

For a system of superballs, the packing fraction is the fraction of space
occupied by the particles, $\phi=\rho v_{sb}$, where $\rho$ is the number
density, $v_{sb}$ is the volume of a superball, 
\begin{equation} 
v_{sb} = \frac{2}{q^2}B\left(\frac{1}{2q},\frac{2q+1}{2q}\right)
B\left(\frac{1}{2q},\frac{q+1}{q}\right), 
\end{equation} 
$B(x,y) = \Gamma(x)\Gamma(y)/\Gamma(x+y)$, and $\Gamma(x)$ is the Euler gamma
function \cite{jiao2009ops}. In this study, we do not consider the family of
superballs for which $q<1$. Superballs are well-suited to study the effect of
the curvature of edges and corners. Although experimentalists have the ability
to control shape and symmetry of colloidal particles, controlling the curvature
may be a larger challenge. Scanning electron micrographs of nanoparticles reveal
that edges and corners are not necessarily sharp like perfect hard polyhedra
\cite{tao2008shape}.  Therefore, understanding the effects of curvature of hard
particles on the phase diagram is not only of fundamental interest, but also of
practical importance.
  
In this study, we detail the phase diagram of hard superball systems as a
function of $q$ and $\phi$ using molecular dynamics. To characterize the
equilibrium phase behavior, it is useful to start a system in an equilibrium
state and allow the particles to grow or contract (or equivalently, allow the
simulation cell to shrink or expand).  The ideal gas is a suitable initial
condition to study low-density phases. Fast growth rates applied to liquids of
hard superballs generate nonequilibirum, randomly jammed packings with novel
characteristics \cite{delaney2010packing, jiao2010dfa}. However, slow growth
rates can also result in nonequilibrium glasses and defective crystals and can
limit access to complex, high-density equilibrium phases. 
 
The densest-packing arrangement is a suitable initial condition for high-density
phases since this arrangement minimizes free energy for a hard-particle system.
This study was motivated by and made possible by the recent development of
optimal packings for the entire family of superballs \cite{elkies1991packing,
jiao2009ops}.  Jiao {\it et al}.\ determined the densest known, and likely
optimal, packings of superballs \cite{jiao2009ops} and superdisks, the
two-dimensional analog \cite{jiao2008ops}. The optimal packings of spheres was
proved rigorously only recently \cite{hales2005proof}, and advances in the
densest-known packings of aspherical particles, including ellipsoids
\cite{donev2004unusually} and the Platonic and Archimedean solids
\cite{torquato2009dense, torquato2009dense2}, will allow researchers to explore
the entire phase diagram of these hard particles.  

The low-density and liquid crystalline phases of many hard-particle systems have
received significant research attention including those for cylinders
\cite{blaak1999cylinders}, spherocylinders \cite{veerman1990phase}, ellipsoids
\cite{straley1974ordered, frenkel1984phase, frenkel1985her, camp1997phase}, cut
spheres \cite{veerman1992phase, duncan2009theory}, tetragonal parallelepipeds
\cite{john2008pbc}, and parallel superellipsoids, a perturbation from ellipsoids
to cylinders \cite{martinez2008nlc}.  However, for many of these shapes, the
optimal packings are yet to be identified ({\it e.g.} ellipsoids
\cite{donev2004unusually}), and therefore, exploring the high-density
equilibrium phases remains challenging.

In the following, we examine the liquid equation of state (EOS), structure of
the liquid phase, and the virial expansion.  For ordered, high-density phases of
superballs, we examine the crystal branch EOS and use suitable order parameters
and correlation functions to characterize the entropy-driven phase transitions
and cubatic ordering.  We find that
\begin{itemize}
\item Excluded volume effects are dominated by edges and corners (Sec.\
\ref{sec:liquid}),
\item There exists a phase transition along the ordered, high-density branch of
the EOS that is associated with changes in long-ranged orientational order
(Sec.\ \ref{sec:crystal}), 
\item The extent of orientational order increases with $q$ at all densities
(Sec. \ref{sec:structure}), and
\item Fixed system boundaries can produce apparent phase transitions in the
ordered, high-density systems. (Appendix \ref{sec:appendix}) 
\end{itemize}
The remainder of this paper is organized as follows. We introduce the deformable
box molecular dynamics methodology, order parameters, and correlation function
in Sec.\ \ref{sec:methods} and review the equations of state for the hard-sphere
and hard-cube systems in Sec.\ \ref{sec:reference}.  In Sec.\ \ref{sec:liquid},
we compare the virial expansion and approximate equations of state to simulation
data. In Sec.\ \ref{sec:crystal}, we present the EOS for the ordered,
high-density and crystal phases as generated with deforming box MD simulations
and show the onset of cubatic ordering as a function of $q$.  The freezing
transitions are examined in Sec.\ \ref{sec:freeze} while the phase diagram is
illustrated in Sec.\ \ref{sec:phase}.  The structural characteristics are
investigated in Sec.\ \ref{sec:structure} and discussions are provided in Sec.\
\ref{sec:discuss}. In Appendix \ref{sec:appendix}, we discuss the role of system
geometry on the isotropy of the internal stresses, particularly how fixed
boundaries can induce apparent, misleading phase transitions.

\section{Methods}
\label{sec:methods}
 
\subsection{Molecular Dynamics}
Simulation studies of the phase behavior of nonspherical particles typically use
Monte Carlo simulation methods, primarily due to the simplicity of
implementation. One drawback of Monte Carlo methods is the difficulty in
achieving collective motion among the particles, which is often an important
characteristic in systems of anisotropic phases. To generate the pressure
equations of state for cube-like superballs, we use the
Donev-Torquato-Stillinger (DTS) molecular dynamics algorithm
\cite{donev2005neighbor1, donev2005neighbor2,donev2006thesis}.  The DTS
algorithm generalizes the Lubachevsky-Stillinger sphere-packing algorithm 
\cite{lubachevsky1990geometric} to nonspherical, convex particles including
superballs and ellipsoids.  In this algorithm, particles are allowed to grow at
a specified nondimensional rate $\gamma$ (or contract for $\gamma < 0$), or
equivalently, the system is compressed (or expanded). Contact between particles
is predicted using generalized overlap potentials. 

The algorithm allows for shape deformation of the boundary using an approach
similar to Parrinello-Rahman MD \cite{parrinello1981polymorphic}. In
Parrinello-Rahman MD, the ``coordinates'' of
the simulation cell are continuously driven by the internal stresses of system
which are directly related to particle interactions.  However, in an
event-driven simulation of hard particles, particles only interact upon contact
and cannot directly interact with the cell. In this paper, a
Parrinello-Rahman-like algorithm is employed where the ``velocities'' of the
lattice vectors are 
updated after a certain number of collisions based upon the anisotropy of stress
tensor \cite{donev2005neighbor1, donev2006thesis}.   The mass assigned to the
cell is equal to that of the total mass of the particles.

In the work presented here, we have verified that the pressure tensor remains
isotropic on average when employing the deforming box algorithm. Although this
algorithm may not rigorously sample an isostress or constant-pressure ensemble,
it is a reasonable approximation. We note that simulations using fixed
boundaries exhibited pressure tensors that were anisotropic and gave rise to
nonequilibrium phase behavior as detailed in the Appendix.

We limit our study to $q\leq4$ since the algorithm is numerically unstable for
$q>4.0$ \cite{donev2006thesis}. Periodic boundary conditions were employed.  The
reduced pressure is defined as $Z = p/\rho k_BT$, where $p$ is the system's
pressure, $\rho$ is the number density, and $k_BT$ is the usual energy scale for
hard-particle systems. Particles are of unit ``diameter,'' the
surface-to-surface distance of a chord along one principal axis ({\it i.e.} the
shortest chord).

To obtain the liquid EOS, particles were placed randomly in a low-density
configuration inside a cubic box. They were given random linear and angular
velocities and allowed to grow at a specified rate $\gamma$ until the system
reached a defined pressure.  For the crystal branch, particles were initialized
in a slightly expanded form of the densest lattice configuration, with the
number of particles $N$ chosen to be commensurate with the lattice, assigned
random linear and angular velocities, and simulated using a contraction rate
$\gamma <0$. 

The densest-known packings of cube-like superballs occur in one of two families
of Bravais lattices, denoted as $\mathbb{C}_0$ and $\mathbb{C}_1$
\cite{jiao2009ops}.  For $1 \leq q \leq 1.1509$, the densest packings of
superballs are achieved with the $\mathbb{C}_0$ lattice, a perturbation of the
FCC lattice.  For superballs with $q \geq 1.1509$, the $\mathbb{C}_1$ lattice, a
deformation of the simple cubic lattice, represents the densest arrangement.
Since the MD algorithm is slow at high densities due to the high frequency of
particle collisions, the initial crystal configurations were unsaturated,
typically near 80\% of the maximum possible packing fraction.

Growth rates in the range $10^{-6} \leq |\gamma| \leq 10^{-3}$ were utilized.
The simulation data for spheres was compared to widely-accepted data and we find
that equilibrium was well approximated with $|\gamma| \leq 10^{-5}$.  Obtaining
a full sweep of the density at $|\gamma| = 10^{-5}$ required over two weeks of
computation time with 1000 particles, and therefore significantly slower growth
rates over the entire density range were not practical.  

We more closely examined parts of the phase diagram in which phase transitions
were evident by running the algorithm with slower rates using near-equilibrium
configurations as the initial conditions.  In some cases, we averaged over
constant-density ($\gamma = 0$) MD trajectories consisting of nearly $10^8$
collisions per particle.  We find that the phase transitions in the $\gamma=0$
cases are slightly sharper than those in cases using slow growth or contraction
rates but generally occur at the same densities.  Growth rates of $\gamma >
10^{-3}$ usually produced jammed, metastable structures. These are explored in
separate studies \cite{delaney2010packing, jiao2010dfa}.  For each $q$, we
obtained multiple, independent sweeps of the density for both the liquid and
solid branches to examine the variability of the results.  The results presented
in this paper were obtained using 1000 particles. We have varied the system size
between $N=216$ to 1728.

It is important to point out that no simulation method can effectively determine
rare events. Even detailed free-energy calculations may encounter the inability
to find states associated with rare events.  Unfortunately, hard-particle MD
requires serial calculations and advances in the parallelization of these
algorithms are required to produce longer trajectories. Until then, these
algorithms are the most efficient use of computer resources and can complement
Monte Carlo methods that lack the ability to capture dynamics of collective
motion.

\subsection{Quantifying Order}

Superballs have three equivalent principal, mutually orthogonal axes labeled $A,
B$ and $C$.  For each axis $j = A, B$, or $C$ of superball $i$, there is an
associated unit vector ${\bf u}_{ij} = [u_{ij,x}, u_{ij,y}, u_{ij,z}]$. For a
nematic-forming system, there is at least one ``director,'' ${\bf n} = [n_x,
n_y, n_z]$, which represents the most aligned direction in a system.  For
sufficiently anisotropic (large $q$) superballs, one might expect systems to
have at least one nematic direction in a low-density phase and three orthogonal
directions in a crystal phase. 

Order parameters are useful to characterize the local and global order in a
system of particles and several have been employed for particles with cubic
symmetry \cite{john2008pbc}.  We find that the nematic and cubatic order
parameters, $S_{2,j}$ and $S_4$ respectively, are the most useful scalar metrics
for quantifying orientational order. For cubatic ordering, a nematic order
parameter can describe ordering in each direction $A, B$ and $C$. The nematic
order parameter for a particular set of axes $j$, $S_{2,j}$  is defined as
\begin{equation}
\label{eq:nem}
  S_{2,j} = \max_{{\bf n}_j} \frac{1}{N}\sum_{i}\left(\frac{3}{2}|{\bf u}_{ij}
\cdot {\bf n}_j|^2-\frac{1}{2}\right)
\end{equation}
where $N$ is the number of particles, ${\bf u}_{ij}$ is a set of particle axes,
and ${\bf n}_j$ is the director for direction $j$. The solution to Equation
(\ref{eq:nem}) can be found by solving the eigenvalue problem ${\bf An}_j =
\lambda{\bf n}_j$ where 
\begin{equation}
 A_{l,m} = \frac{3}{2N} \sum_{i}u_{ij,l}u_{ij,m}-\frac{1}{2}\delta_{l,m},
\end{equation}
where $\delta$ is the Kronecker delta function.  $S_{2,j}$ is the maximum
eigenvalue $\lambda_{max}$ and the nematic director vector ${\bf n}_j$ is the
eigenvector associated with $\lambda_{max}$.  

Since all principal axes of a superball are equivalent, we must ``relabel'' the
particle axes prior to calculating $S_{2,j}$ in order to obtain meaningful data.
A set of three mutually orthogonal unit vectors is chosen as a reference,
typically the principal axes of one randomly chosen particle or the standard
laboratory axes. For each particle, we relabel the particle axes based on the
best alignment with the reference system.  For example, we identify the axis of
each superball that is best aligned with the $[1,0,0]$  vector, label these axes
as $A$ axes, then continue with the $[0,1,0]$ vector and $B$ axes. The remaining
axes are labeled as $C$ axes. A schematic of this procedure is shown in Ref.\
\cite{john2004clc}.  The relabeling scheme introduces artificial correlations so
that $S_{2,j}$ in an isotropic system is approximately 0.55.  For perfect
cubatic ordering, $S_{2,j}=1$ in each of the three orthogonal directions.  Here,
we report $S_2$ as the maximum of the $S_{2,j}$'s since the $S_{2,j}$'s were
nearly equal to each other in all of the cases considered.  This suggests that
we encountered isotropic and cubatic phases and not phases with strict uniaxial
or biaxial order.
 
The cubatic order parameter $S_4$ is a more appropriate scalar metric for
ordering in three orthogonal directions and is defined as 
\begin{equation} S_4 =
\max_{\bf n} \frac{1}{14N} \sum_{i,j} \left(35|{\bf u}_{ij} \cdot {\bf n}|^4 -
30|{\bf u}_{ij} \cdot {\bf n}|^2 +3\right). 
\end{equation} 
Here, ${\bf n}$ is a unit director for which $S_4$ is maximized. The prefactor
of $1/14N$ arises from the accounting for the $3N$ principal axes and
normalizing $S_4$ to unity for perfect alignment. This can be formulated into a
eigentensor problem as in Ref.\ \cite{duncan2009theory}. However, we use an
approximate solution. We choose the maximum $S_4$ from a large set of trial
directors ${\bf n}$.  Here, we take the set of trial directors to be the set of
all particle axes ${\bf u}_{ij}$, providing $3N$ trial directors. We report the
maximum $S_4$ from this set of trial directors. For perfect cubatic ordering,
$S_4=1$, and for a system with no long-range cubatic order, $S_4=0$. 

We use an orientational correlation function $G_4(r)$ to measure the mutual
alignment of particles as a function of distance $r$ between particle centers.
As a specific instance of a general class of orientational correlation functions
\cite{veerman1992phase}, $G_4(r)$ is defined as
\begin{equation}
\label{eq:ocf}
G_4(r) = \frac{3}{14} \left<35 \left[{\bf u}_{aj}(0)\cdot {\bf
u}_{bj}(r)\right]^4 - 30 \left[ {\bf u}_{aj}(0)\cdot {\bf u}_{bj}(r) \right]^2
+3\right>,
\end{equation}
where the $<\cdots>$ denotes the average over all axes $j$ and particle pairs
$a$ and $b$. The prefactor of $3/14$ accounts for all nine combinations of axes
between particle pairs which is similar to the normalization of $S_4$. In the
limit that $r\rightarrow\infty$, $G_4(r)$ approaches $S_4^2$.  This function
allows us to determine local correlations whereas $S_4$ determines global order.

The local alignment between ``neighbor'' particles can be characterized by the
angular distribution function $f(\theta)$ defined for $0^\circ \leq \theta <
180^\circ$. The probability of finding two neighbor particles whose axes are
aligned at an angle between $\theta$ and $\theta + d\theta$ is given by
$f(\theta)d\theta$.  We consider two particles to be neighbors if they are
separated by less than 1.35 diameters, since for most particles, this is between
the first and second neighbor shells of the associated crystal.  However,
$f(\theta)$ is relatively unaffected by small variations of the cutoff radius.
For the perfect crystal of cube-like particles, $f(\theta) =
\frac{1}{3}\delta(\theta) + \frac{2}{3}\delta(\theta-90^\circ)$, where
$\delta(\theta)$ is the Dirac delta function.  At equilibrium, $f(\theta$) is
symmetric about $\theta=90^\circ$. 
 
The order parameters derived from the spherical harmonics
\cite{steinhardt1983boo} were calculated but do not add significantly to our
analysis. The use of a different reference crystal for each $q$ precludes our
ability to compare superballs with different $q$.  Also, changes in these order
parameters as a function of $\phi$ closely matched changes in $S_2$ and $S_4$.

To examine translational order, we use the radial distribution function
$g_2(r)$, which is the normalized pair density distribution function such that
for a disordered system it tends to unity for large pair separation $r$. In
addition, one-dimensional particle distribution functions were calculated in
each of the three associated one-dimensional directions of the crystal. When
simulating the melting of the crystal, these particle distribution functions
remained periodic until the crystal melted.  This confirms that
lower-dimensional translational order, such as that found in columnar or smectic
phases, was never present.

\subsection{Virial Coefficients}

The virial expansion for hard particles in terms of reduced pressure $Z$ is
given by
\begin{equation}
 Z = p/\rho kT = 1 + \sum_{i=2}^\infty \left(B_i/v^{i-1} \right)\phi^i, 
\end{equation}
where $p$ is the pressure, $B_i$ is the $i^{th}$ virial coefficient, and $v$ is
the volume of a single particle. For hard, convex particles the second virial
coefficient is known analytically as $B_2 = RS + v$ \cite{isihara1951theory},
where $R$ and $S$ are the radius of mean curvature and surface area of a
particle, respectively.  Evaluating analytic expressions for these quantities is
highly nontrivial, as was the case for ellipsoids \cite{singh1996geometry}, but
numerical calculation of the second and higher virial coefficients is
straightforward given a suitable overlap function. We calculate the first few
virial coefficients using Monte Carlo integration \cite{metropolis1953equation}.
Trial configurations were generated using the method of Ree and Hoover
\cite{ree1964fifth}. For $B_2$, $1.5\times 10^6$ random trial configurations
were used.  For $B_3$, random configurations were generated until $2\times10^6$
configurations satisfied the condition that particle 1 overlaps particle 2, 2
overlaps 3, and 3 overlaps 1.  For $B_4$, random configurations were produced
until $2\times10^6$ configurations satisfied the condition that particle 1
overlaps 2, 2 overlaps 3, 3 overlaps 4, and 4 overlaps 1. The standard
deviations of ten subaverages were less than 0.5\% of the virial coefficient.
The algorithm was tested against known results for hard ellipsoids
\cite{vega1997virial}. We have calculated hard-cube virial coefficients to a
higher accuracy than in Ref.\ \cite{nezbeda1984possible} using the separating
axis theorem to check for overlaps.

\section{Reference Systems: Hard Spheres and Hard Cubes}
\label{sec:reference}

\begin{figure}
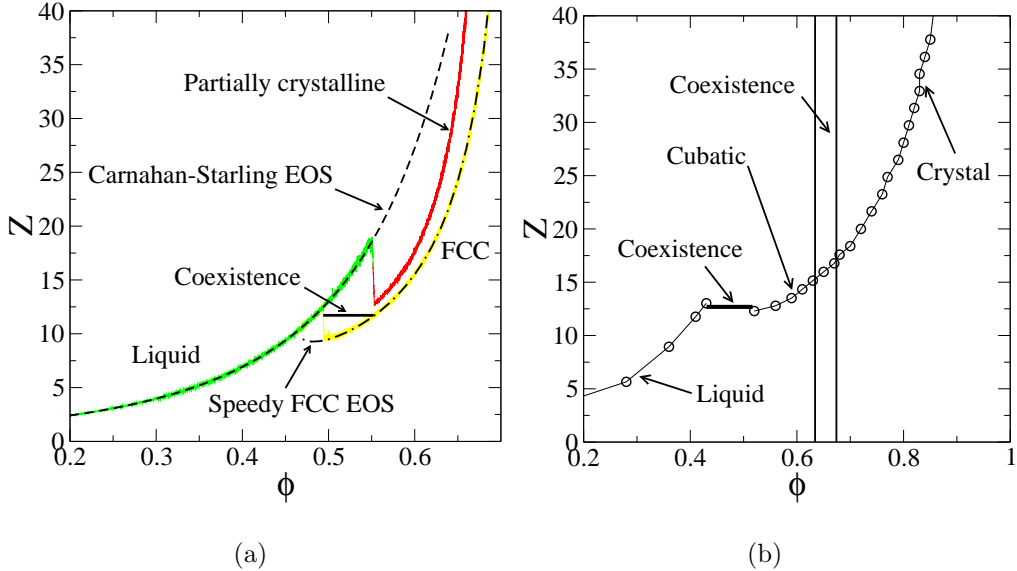

\subfigure[]{\label{fig:sphereref} \includegraphics[width=0.4\textwidth, clip=true]{Fig2a.eps}}
\subfigure[]{\label{fig:cuberef}   \includegraphics[width=0.4\textwidth, clip=true]{Fig2b.eps}}
\caption{(Color online) (a) Pressure EOS for hard spheres generated by the DTS
algorithm with $N=1000$. The liquid and partially crystalline branches were
obtained starting from a random initial configuration and $\gamma=10^{-5}$ while
the crystal branch was obtained by initializing the system in FCC arrangement
and $\gamma=-10^{-5}$. The coexistence line is that of Refs.\
\cite{de2008estimating}. (b) Pressure equation of state and coexistence line for
freely rotating hard cubes from Refs.\ \cite{john2008pbc, footnote}.} 
\end{figure}

Since superballs interpolate between a sphere and cube for $q$ ranging from
unity to infinity, we review the phase behavior of these reference systems. The
hard-sphere liquid and crystal have been well-characterized by simulation and
theoretical treatments.  The hard-sphere EOS as generated by the DTS algorithm
with $\gamma = \pm 10^{-5}$ and $q=1$ is shown in Fig.\ \ref{fig:sphereref}.
Although the DTS algorithm does not achieve true equilibrium, which requires
$\gamma=0$ for long MD trajectories, the slow growth rate approximates
equilibrium well. The DTS algorithm correctly produces the well-known and
widely-accepted Carnahan-Starling EOS for liquids \cite{carnahan1969equation}
and the empirically-derived Speedy EOS for the FCC crystal
\cite{speedy1998pressure}.

The freezing and melting points of a hard-sphere system are $\phi = 0.490$ and
$\phi = 0.545$, respectively, with a coexistence pressure $Z=11.48$
\cite{de2008estimating}.  In Fig.\ \ref{fig:sphereref}, one can see that the DTS
algorithm produces a first-order phase transition at $\phi=0.551$ to a partially
crystalline system when starting the system in a random initial configuration
and allowing the particles to grow slowly at $\gamma = 10^{-5}$. This packing
fraction is repeatable over independent runs and consistently occurred between
$\phi=0.545$ and 0.553.  Halving the growth rate to $\gamma = 5\times 10^{-6}$
(not shown in figure) resulted in a freezing event at $\phi=0.547$.  Starting in
the FCC crystal arrangement and using $\gamma = -10^{-5}$, the simulation data
traces the Speedy EOS and the system shows a first-order transition at $\phi =
0.495$.  The density at which the phase transitions occurred showed little
variability, occurring between $\phi=0.496$ and 0.498 across multiple runs.
Halving the rate to $\gamma = -5\times 10^{-6}$ yielded a transition at $\phi =
0.496$.  The algorithm evidently is appropriate for determining the densities at
which phase transitions occur and identifying coexistence regions. The DTS
algorithm, however, cannot explicitly identify the coexistence pressure, which
requires extensive free-energy calculations. An equal-area construction of the
pressure-volume equation of state may provide an approximation to the
coexistence pressure.
 
There has been considerably less attention devoted the hard-cube system.
However, several studies have elucidated the EOS, phase transitions, and
orientational ordering. Parallel hard cubes undergo a continuous melting
transition from a crystal to a liquid \cite{jagla1998mhc, groh2001closer}. The
EOS for parallel hard cubes has several gentle curvature changes but no evidence
of a first-order phase transition \cite{hoover2009single}.

However, we are interested in the EOS and phase transitions in systems of
freely-rotating hard cubes.  John {\it et al}.\ obtained the phase diagram for
freely rotating hard cubes by Monte Carlo methods \cite{john2008pbc} and their
EOS is shown in Fig.\ \ref{fig:cuberef}.  A first-order melting transition was
first calculated to be between $\phi = 0.45$ and 0.52 \cite{jagla1998mhc}.  More
recent studies reveal a narrower coexistence region between an isotropic and a
cubatic phase to be in the region $0.437 \leq \phi \leq 0.495$
\cite{john2008pbc}.  More interestingly, John and coworkers suggest that there
is a cubatic-crystalline transition between $\phi = 0.634$ and $0.674$. However,
the nature of the transition was not well-characterized and necessitated
free-energy calculations for verification \cite{john2008pbc}. The DTS algorithm
is numerically unstable for superballs with large $q$ and therefore, these Monte
Carlo calculations are the best available reference data.

\section{Equations of State}

\subsection{Isotropic Liquid Phase}
\label{sec:liquid}

\begin{figure}
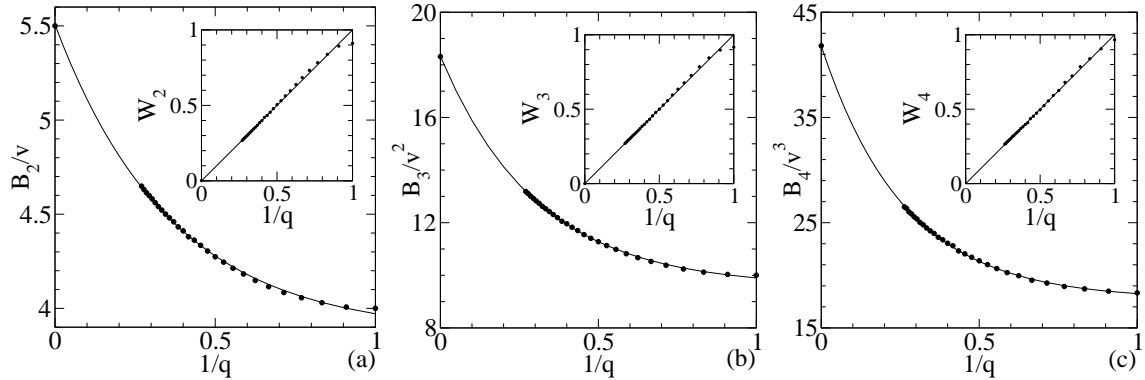

\includegraphics[width=0.3\textwidth, clip=true]{Fig3a.eps}
\includegraphics[width=0.3\textwidth, clip=true]{Fig3b.eps}
\includegraphics[width=0.3\textwidth, clip=true]{Fig3c.eps}
\caption{The (a) second, (b) third, and (c) fourth virial coefficients for hard
superballs fit to the equation $\frac{B_i}{v^{i-1}} = a_i \exp(-b_i/q) + c_i$.
The inset shows $W_i =
-\frac{1}{b_i}\ln\left[\left(\frac{B_i}{v^{i-1}}-c_i)\right/a_i\right]$. The
table of values for the fitting coefficients is provided in the text.} 
\label{fig:virials.fit}
\end{figure}

The low-order virial coefficients describe the low-density EOS of a system.  The
second, third, and fourth virial coefficients, shown in Fig.\
\ref{fig:virials.fit}, monotonically increase with the deformation parameter
$q$. This is expected for $B_2$ and $B_3$ given that the cluster integrals
involved in the calculations effectively measure excluded volume, and a larger
particle generally yields a larger excluded volume.  The effects on $B_4$ and
higher-order coefficients are not intuitively obvious, since these coefficients
are not necessarily strictly positive for hard particles. The virial
coefficients are closely approximated by an exponential equation, 
\begin{equation}
\frac{B_i}{v^{i-1}} = a_i \exp(-b_i/q) + c_i.
\end{equation}
The values of the parameters obtained by nonlinear least-squares regression are
given in Table \ref{tab:virials}.  These parameters were the best obtained,
though they are not guaranteed to be 
 optimal due to the nonlinearity of the fit.  The insets in Fig.\
\ref{fig:virials.fit} plot 
\begin{equation}
 W_i = -\frac{1}{b_i}\ln\left[\left(\frac{B_i}{v^{i-1}}-c_i)\right/a_i\right]
\end{equation}
versus $1/q$ to show that the scaling of $W_i$ is approximately linear with
$1/q$, except near the sphere point $q=1$ where the fit tends to degrade.

A significant portion of the excluded volume evidently arises from the
particles' edges and corners. Sharpening of edges and corners increases the
excluded volume in an isotropic phase faster than it increases the actual volume
of a particle.  When visually comparing a superball with a moderately large $q$
value, say $q=4$, to the perfect cube (see Fig.\ \ref{fig:superballs}), one
might surmise that because the two particles have such similar appearances, the
two particles would have similar behavior in the liquid phase. However, when
comparing $B_2/v$, the relative excluded volume, for these two particle shapes,
the superball with $q=4$ is closer to a sphere than to a cube. The edges and
corners evidently are dominant features contributing to excluded volume effects.

\begin{table}
\caption{\label{tab:virials} Coefficients of fit for $B_i/v^{i-1} = a_i
\exp(-b_i/q) + c_i$}
\begin{ruledtabular}
\begin{tabular}{lccc}
   & $B_2/v $  & $B_3/v^2$ & $B_4/v^3$ \\ 
\hline
$a_i$ & 1.638 & 8.759 & 24.077 \\
$b_i$ & 2.770 & 3.262 &  3.794 \\
$c_i$ & 3.869 & 9.564 & 17.715 \\
\end{tabular}
\end{ruledtabular}
\end{table}

\begin{figure}
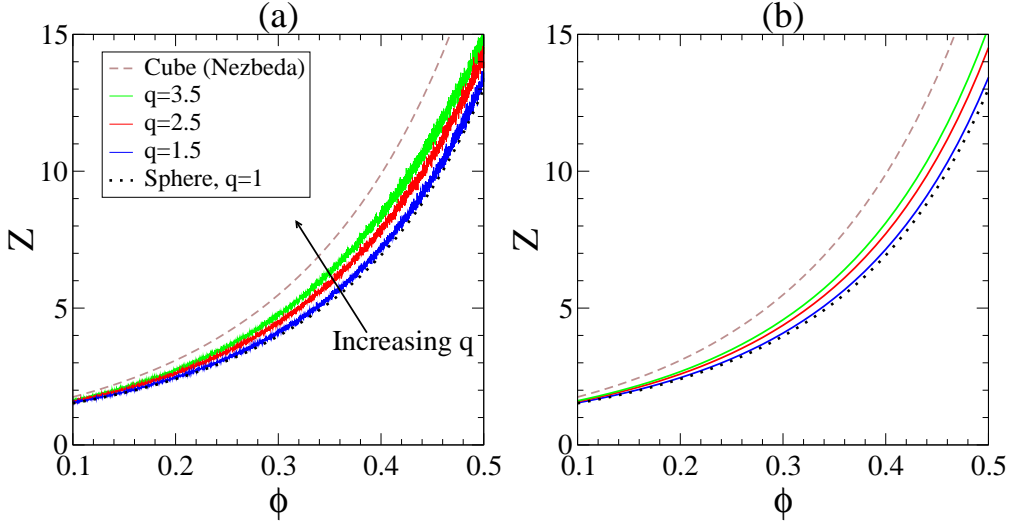

\includegraphics[width=0.4\textwidth, clip=true]{Fig4a.eps}
\includegraphics[width=0.4\textwidth, clip=true]{Fig4b.eps}
\caption{(Color online) Liquid equation of state for $q=1.5$, 2.5, and 3.5 from
(a) simulation and (b) the Nezbeda EOS.  When overlaid, the Nezbeda EOS is
accurate for low densities but accelerates its divergence at higher densities,
typically for $\phi>0.25$ and for $q>2.5$.  } 
\label{fig:liq.sim}
\end{figure}

Fig.\ \ref{fig:liq.sim}a shows the simulation data for several $q$ values for
low densities.  We compare our simulation results to the Nezbeda EOS
\cite{nezbeda1979colln, boublik1981equation}, a modification of the
Carnahan-Starling EOS \cite{carnahan1969equation} for convex hard particles.
Using only the nonsphericity parameter $a = RS/v$, the Nezbeda EOS is given as 
\begin{equation}
Z = \frac{1}{1-\phi} + \frac{3a\phi}{(1-\phi)^2} + \frac{3a^2\phi^2 -
a(6a-5)\phi^3}{(1-\phi)^3}.
\end{equation}
The Nezbeda curve, Fig.\ \ref{fig:liq.sim}b, follows simulation data of
superballs along the entire liquid branch for $q<2.5$. For $q\geq 2.5$, the
Nezbeda curve follows the simulation data at low densities, slightly
underestimates pressures at moderate liquid densities, and slightly
overestimates pressures at higher densities. For example, with $q=2.5$, the
Nezbeda curve is accurate for $\phi \leq 0.20$, underestimates the pressure for
$0.20< \phi< 0.47$, and overestimates the pressure for $\phi \geq 0.47$. Using
local polynomial fits to the simulation curves, we can compare the pressure
values of the simulation curves to those of the Nezbeda equation of state. For
$q=2.5$, $Z_{sim}-Z_{Nez}$ is less than 0.059 at $\phi=0.2$, while for
$\phi=0.35$, $Z_{sim}-Z_{Nez}$ is less than  0.138.  At $\phi = 0.48$,
$Z_{sim}-Z_{Nez}$ is about -0.086. As shown by the simulation data, superballs
that are visually similar to cubes have pressures midway between that of spheres
and that of cubes, demonstrating the effects of sharp edges and corners on the
low-density EOS.  

\subsection{Ordered, High-Density Phases: Melting}
\label{sec:crystal}
 
\begin{figure}
\includegraphics[width=0.4\textwidth, clip=true]{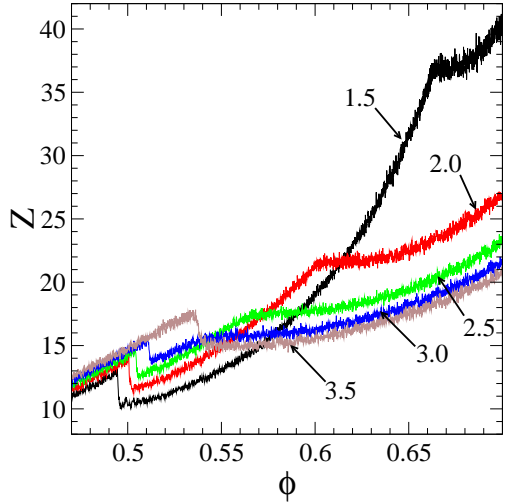}
\caption{(Color online) Crystal branches and melting transitions for various $q$
values obtained using the contraction rate $\gamma = -1\times10^{-5}$. The
transition from the $K$ phase to $Q$ occurs at a lower density for increasing
$q$ until it is absorbed entirely by the $K$ phase.  } 
\label{fig:crys.press}
\end{figure}

To obtain the ordered, high-density equations of state, systems were initialized
in the optimal lattice configuration and a contraction rate of $\gamma =
-10^{-5}$ was applied until the system entered the fluid phase. Figure
\ref{fig:crys.press} illustrates the resulting equations of state for various
values of $q$.  As seen in the figure for $q\leq 3.0$, there exist two apparent
phase transitions, which result in three distinct phases.  We define these
phases, going from highest density to lowest, as the crystal ($K$), cubatic
($Q$), and liquid ($L$) phases. For $q> 3.0$, only one phase transition is
evident from the pressure EOS, separating the $K$ and $L$ phase.  

The $K$ phase is characterized by long-ranged translational and orientational
order, quantifiable by the appropriate order metrics and distribution functions.
 We characterize the $Q$ phase as having a moderate degree of long-ranged
orientational order compared to the crystal phase. Particles are loosely braced
and have an $S_4$ value above that of a dense liquid but less than that of a
crystal, between values of 0.05 and 0.20. In this phase, crystalline
translational order remains. While some have characterized the cubatic phase as
having no long-ranged positional order \cite{blaak1999cylinders}, others also
consider those phases with an ``intermediate degree of translational order'' as
cubatic phases \cite{john2004clc, john2005phase}.  We choose to use the latter
definition of ``cubatic,'' since this was the definition employed for the use of
hard cubes. The $L$ phase lacks long-ranged positional and orientational order. 

\begin{figure}
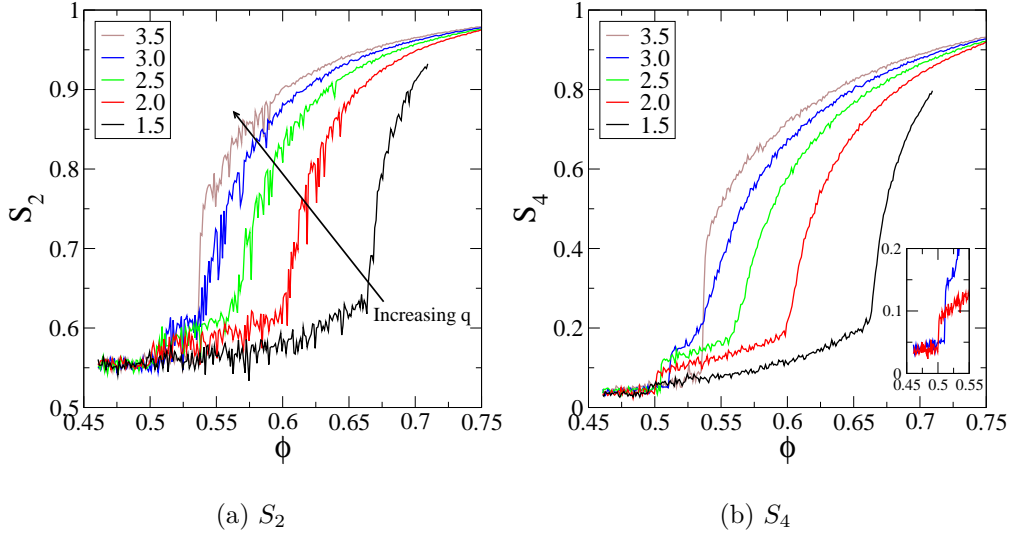

\subfigure[ $\text{ }S_2$
]{\label{fig:nematic}\includegraphics[width=0.4\textwidth,
clip=true]{Fig6a.eps}}
\subfigure[ $\text{ }S_4$
]{\label{fig:cubatic}\includegraphics[width=0.4\textwidth,
clip=true]{Fig6b.eps}}
\caption{(Color online) Nematic and cubatic order parameters $S_2$ and $S_4$ for
$q=1.5$, 2.0, 2.5, 3.0 and 3.5. with $\gamma = -10^{-5}$.  The inset shows the
behavior at the melting transition for $q=2.0$ and 3.0, though the behavior is
similar at this transition for all $q$. The labels on the inset are the same as
the larger graph.}
\label{fig:orderpar}  
\end{figure}

The nematic and cubatic order parameters associated with the ordered,
high-density phases are shown in Fig.\ \ref{fig:orderpar}. In all cases, the
curve for $S_2$ exhibited more noise than the corresponding $S_4$ curve,
possibly due to the ``relabelling'' scheme. Regardless, the behavior of both
order metrics follows closely with the pressure equations of state.  When the
pressure-density curve is discontinuous or shows a change in curvature, the
order parameters exhibited a corresponding change.

Beginning in the densest crystal and moving down along the EOS, the slope of the
reduced pressure is discontinuous at the $K$-$Q$ transition. This transition
point initially occurs at high densities for small $q$. For increasing $q$, this
transition occurs at lower densities and by $q=3.5$, the transition vanishes. 
As seen by the order parameters in Fig.\ \ref{fig:orderpar}, this transition
results in a reduction in cubatic ordering. While the $S_4$-density curve
remains continuous at the transition, the slope does not, similar to the
behavior of the pressure-density curve.  Although the $Q$ phase has less
orientational order than the $K$ phase, there remains significant long-ranged
cubatic order.  At the melting transition from $Q$ to $L$ or $K$ to $L$, the
nematic order parameters are discontinuous as shown in the inset of Fig.\
\ref{fig:orderpar}.

The $K$-$Q$ transition was observed in the EOS for $1.3\leq q\leq 3.0$, however,
we strongly suspect that similar transitions occur for smaller $q$ and high
pressure.  Unfortunately, simulating the system for $q=1.2$ and $\phi>0.72$
proved challenging, both in the stability of the code and achieving
near-equilibrium behavior. Because of the monotonic behavior of the location of
the $K$-$Q$ transition with respect to $q$, we suspect this to continue for $q$
near the sphere point.  For $q>3.0$, we do not observe a $K$-$Q$ transition,
although it is possible that a $K$-$Q$ phase transition exists but is of a
higher order. Higher-order phase transitions were suggested for hard cubes
\cite{john2008pbc}, though in the case of nearly cubic superballs, there is no
discontinuity in the cubatic order parameters aside from the first-order
transition associated with melting.  In addition, we have shown that edges and
corners play an important role in the liquid EOS, and therefore, it is not
unreasonable to expect that the high-density behavior of nearly cubic superballs
($q=4.0$) can deviate from that of hard cubes.

\begin{figure}
\subfigure[ $\text{ }\phi=0.626$]{\label{fig:0.626}
\includegraphics[width=0.40\textwidth, clip=true]{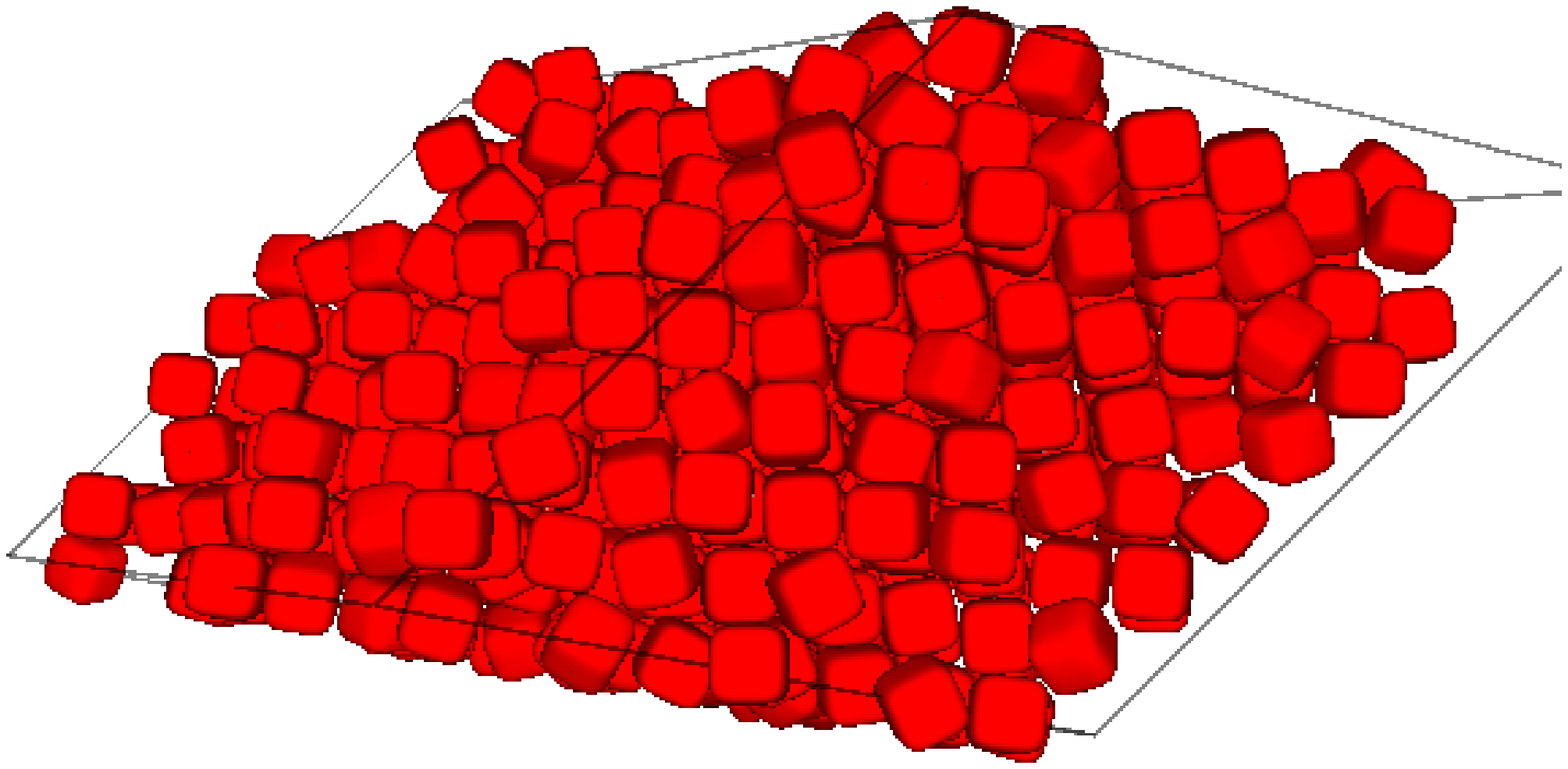}
\includegraphics[width=0.30\textwidth, clip=true]{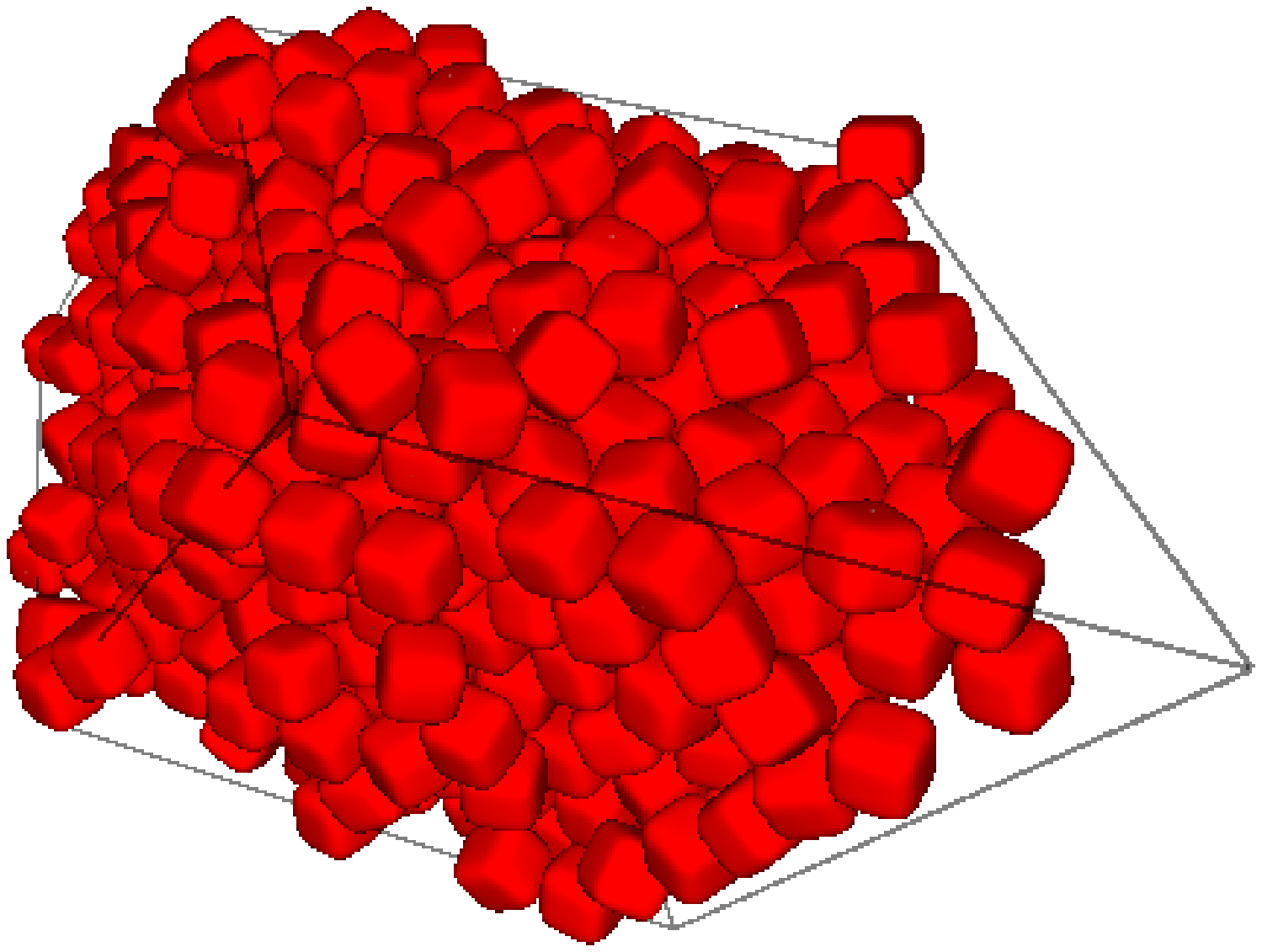}}
\subfigure[ $\text{ }\phi=0.600$]{\label{fig:0.600}
\includegraphics[width=0.40\textwidth, clip=true]{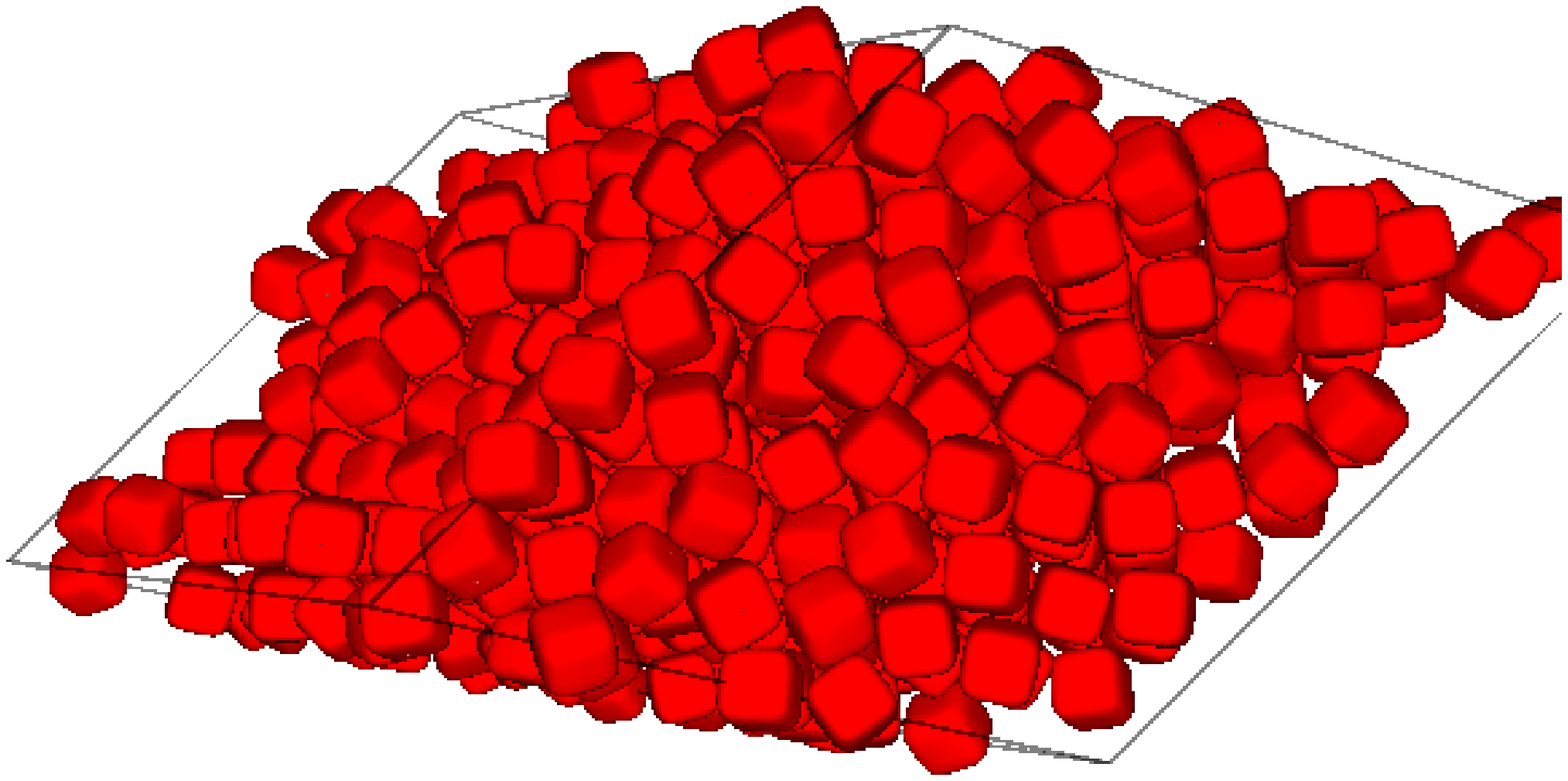}
\includegraphics[width=0.30\textwidth, clip=true]{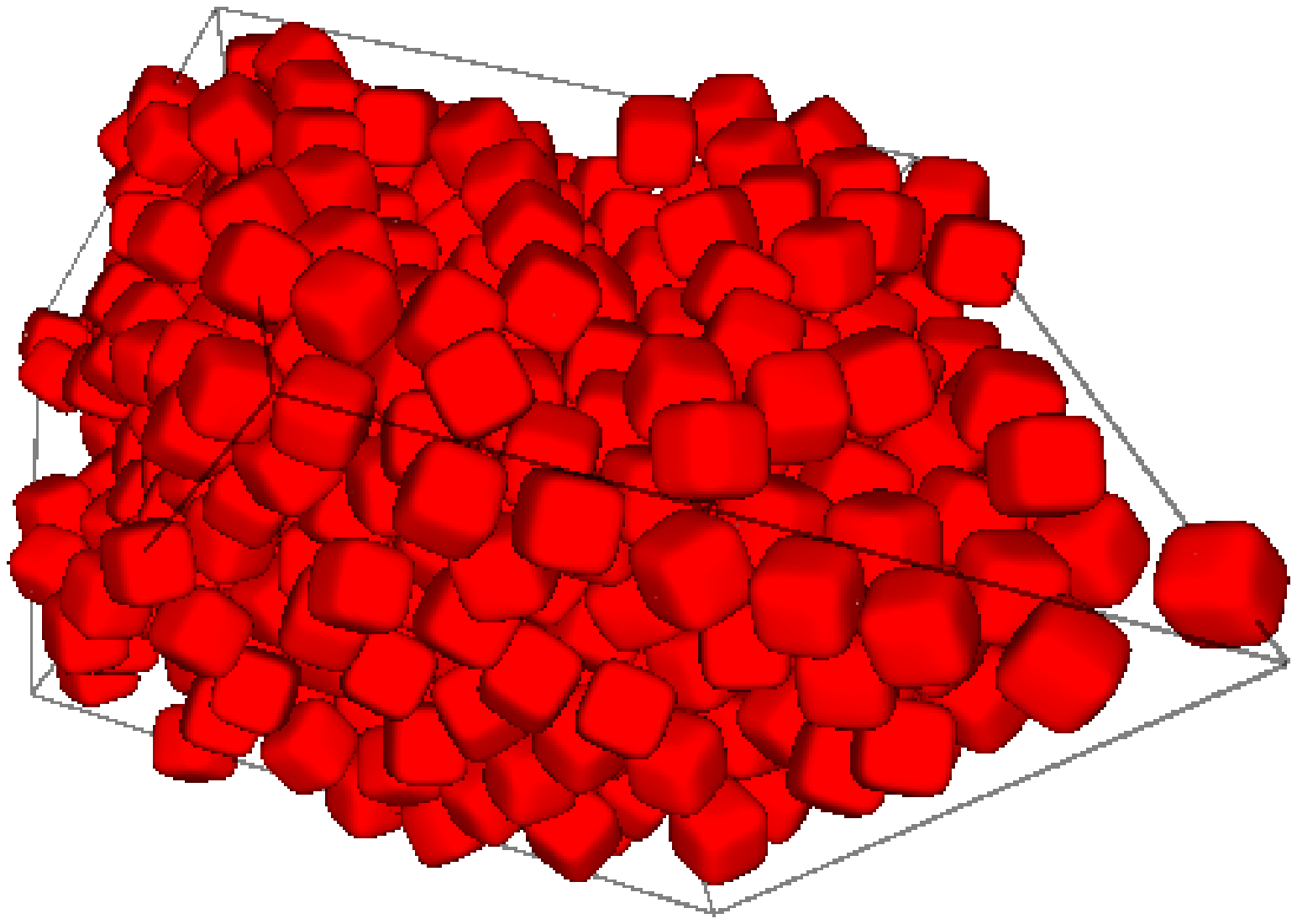}}
\subfigure[ $\text{ }\phi=0.558$]{\label{fig:0.558}
\includegraphics[width=0.40\textwidth, clip=true]{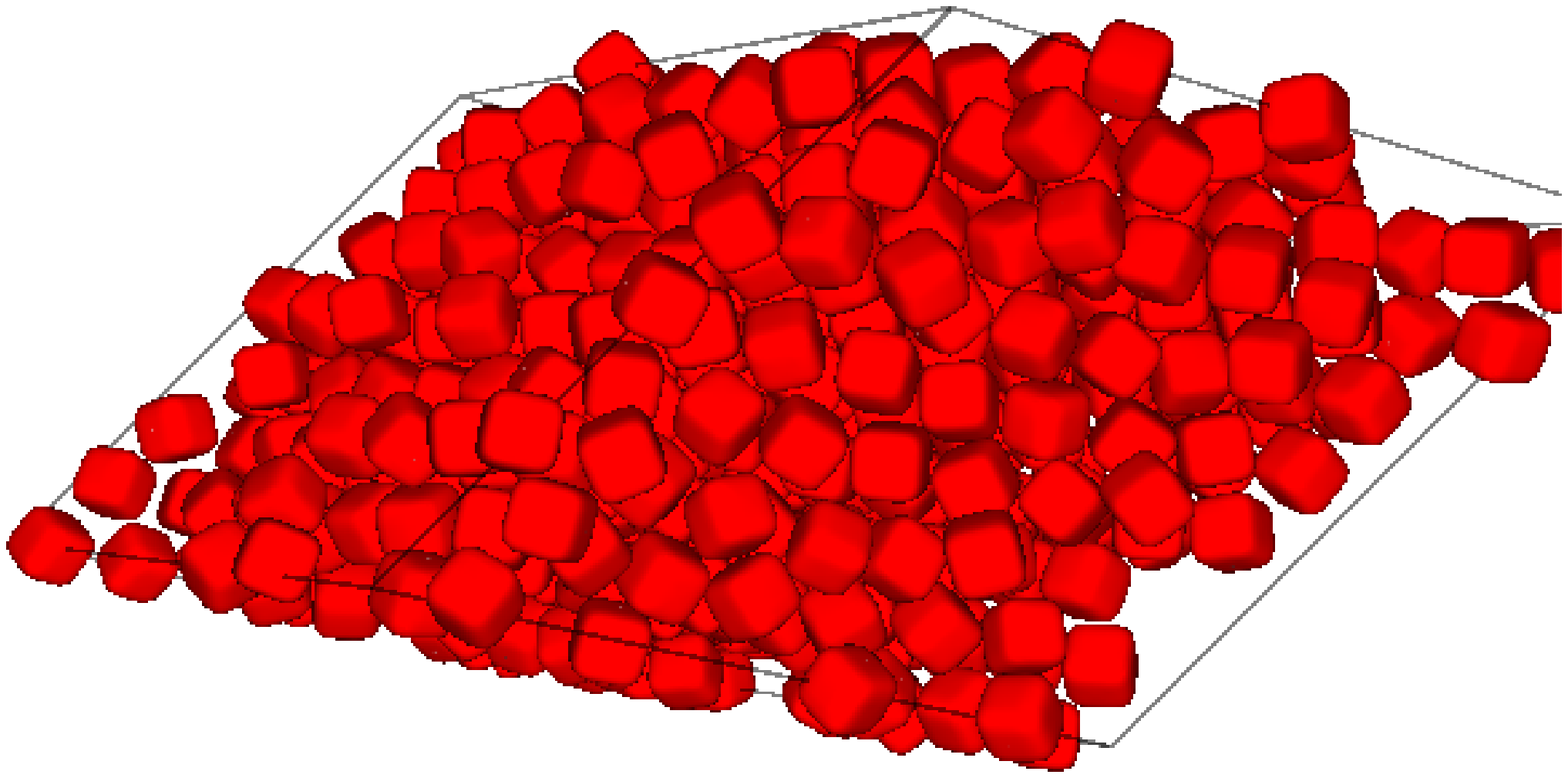}
\includegraphics[width=0.30\textwidth, clip=true]{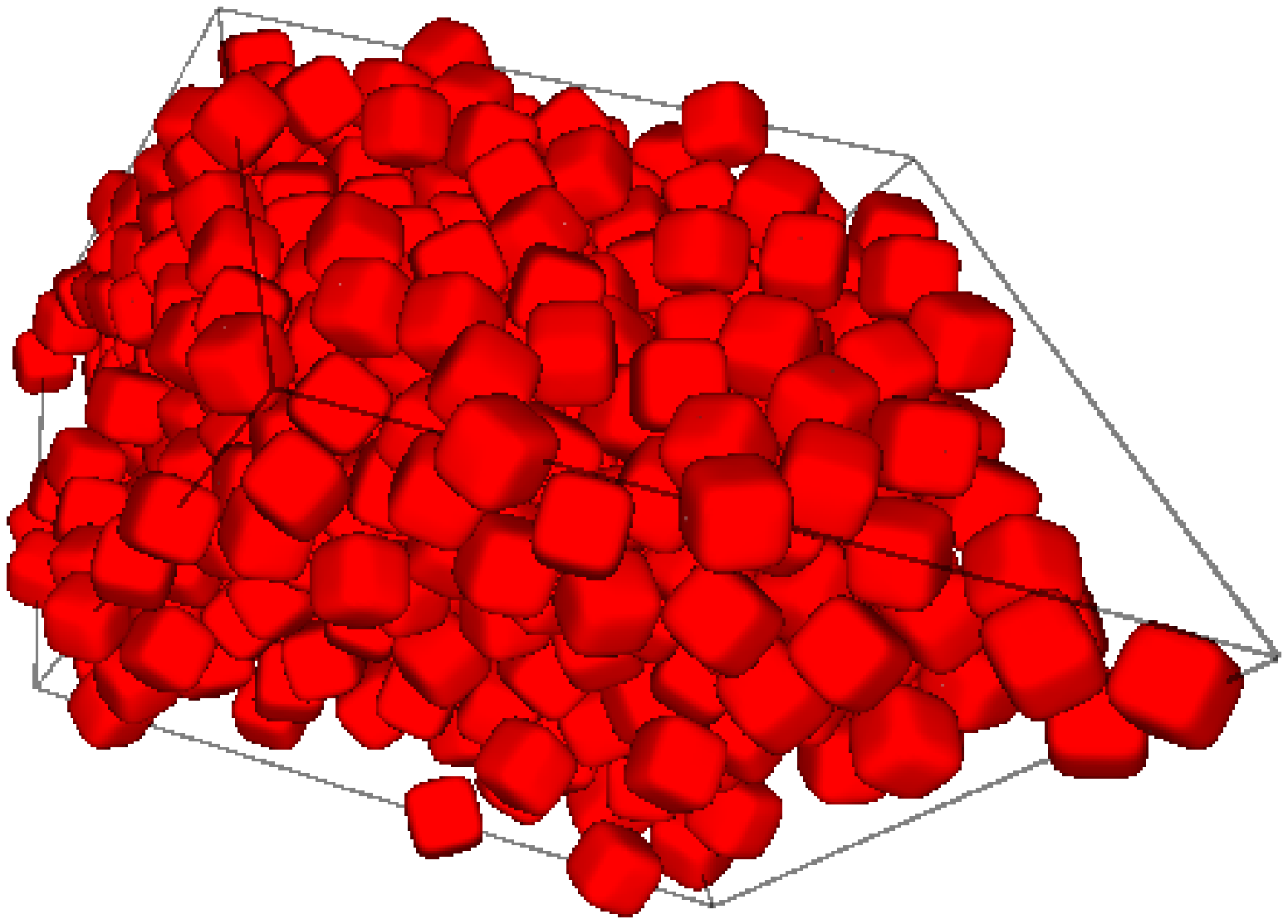}}
\caption{(Color online) Two views of particle configurations from the crystal
branch for $q=2.0$ for a system containing 512 particles.  The dark lines are
the boundaries of the simulation box. One can observe significant translational
order in each. Orientational order is reduced as the density is decreased, but
translational order is maintained.}
\label{fig:image.crys}
\end{figure}

As the density is reduced in the $Q$ phase, the cubatic order parameters
smoothly decreases. At the $Q$-$L$ transition, the pressure jumps while $S_2$
and $S_4$ drop close to the values associated with random rotations.  The
$Q$-$L$ transition is clearly first order and is present for all $q$ tested. 
The density at which melting occurs increases monotonically from $\phi = 0.494$
for hard spheres up to 0.536 for $q=4.0$.  While reducing the density along the
high-density equation of state, the translational order appears to drop
continuously. The peaks in $g_2(r)$ maintain crystal-like characteristics. We
discuss translational order in greater detail in the next section. Figure
\ref{fig:image.crys} shows a representative sample of the system at several
densities in the $K$ and $Q$ phases.  The decreased orientational order
associated with the $Q$ phase is not easily distinguishable from the visual
inspection but is clearly evident by the order parameters.

The decrease in the density associated with the $K$-$Q$ phase transition for
increasing $q$ is attributed to the broadness of the edges and corners of
particles as they relate to particle rotations. 
For each $q$ in the $K$ phase, particles are braced against rotations. Particle
rotations require large local fluctuations in the density to allow for particle
flips. As the relative lattice spacing is increased, the rotational mobility of
the particles increases. With increasing $q$, particles have sharper corners
that are more effective in bracing against rotations for larger density ranges. 
The required local volume necessary for particle rotations increases with more
cube-like shape, and therefore we observe a larger $K$ phase for increasing $q$.

The increased rotational mobility of the $Q$ phase also imparts larger internal
stresses than in the $K$ phase, as observed by the greater slope of the
pressure-density curve in the $Q$ phase than the $K$ phase at the $K$-$Q$
transition.  In Sec.\ \ref{sec:discuss}, we detail some of the tests we
performed to ensure that these results were not related to system-size,
boundary, or kinetic effects.

\subsection{Ordered, High-Density Phases: Freezing}
\label{sec:freeze}

\begin{figure}
\subfigure[ $\text{ }q=1.5$]{\label{fig:p1.5.freeze}
\includegraphics[width=0.35\textwidth, clip=true]{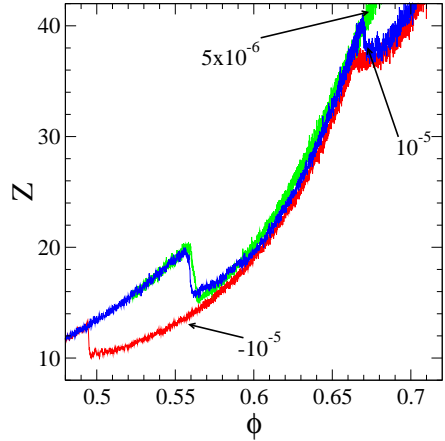}} \\
\subfigure[ $\text{ }q=2.0$]{\label{fig:p2.5.freeze}
\includegraphics[width=0.35\textwidth, clip=true]{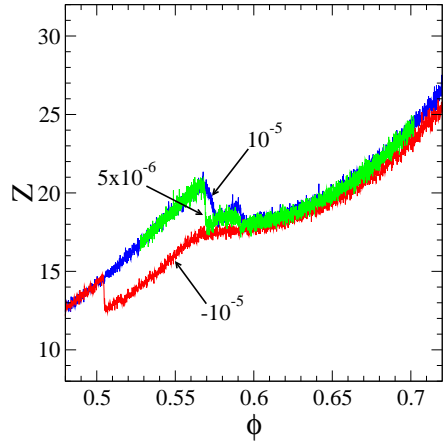}}
\caption{(Color online) Freezing transition for $q=1.5$ and 2.5 and $\gamma =
10^{-5}$ (blue), $5\times 10^{-5}$ (green), and $-10^{-5}$ (red). In all cases
tested, a freezing transition resulted in crystals with vacancies and/or grain
boundaries.} 
\label{fig:crys.press.freeze}
\end{figure}

The freezing transitions for superballs were examined by applying a slow growth
rate to particles initialized in a low-density liquid. Figure
\ref{fig:crys.press.freeze} shows the pressure as a function of packing fraction
for two values of $q$ along with the crystal branch equation of state. In the
figure, two growth rates are displayed, $\gamma = 10^{-5}$ and $5\times10^{-6}$.
  We find that upon freezing most systems order into a partially crystalline
structure representative of a $L$-$Q$ transition. Across this transition, the
pressure drops while the order parameters increase discontinuously. At high
pressures, the particles increase the face-to-face contacts, but the nematic
directors of the separate crystal regions destructively interfere. Therefore,
the order parameters 
cannot achieve values as large as the equilibrium branch at the same density. 
The pressure of the partially crystalline phase is greater than that of the
equilibrium structure, which is evident in Fig.\ \ref{fig:crys.press.freeze}.
These systems have characteristics of translational order of a crystal but have
a large grain boundary or vacancies.  At very high pressures, the grains are
eliminated by the shear deformation of the box.  The structural characteristics
of jammed superballs are detailed in Ref.\ \cite{jiao2010dfa}.

The $Q$-$K$ transition is sensitive to the extent of crystallization occurring
at the $L$-$Q$ transition. In several cases, the system slightly overshoots the
$Q$-$K$ transition while in other cases, the system never shows signs of a
$Q$-$K$ transition. One might expect that using slower growth rates would
alleviate this phenomenon. However, as seen in Fig.\ \ref{fig:p1.5.freeze},
slower growth rates are not necessarily better at achieving the $Q$-$K$
transition and tracing the crystal branch EOS. The density at which the
crystallization from the $L$ phase to the $Q$ phase occurs is relatively
insensitive to the growth rate for the slow rates used in this study. It is of
note that substantially faster growth rates produce amorphous, jammed structures
\cite{jiao2010dfa}. 

For $q\geq2.5$, we find that the initial freezing from a liquid phase occurs at
a density higher than that of the $Q$-$L$ transition along the ordered,
high-density branch which is shown in Fig.\ \ref{fig:p2.5.freeze}.  In these
cases, both growth rates, $10^{-5}$ and $5\times10^{-6}$, resulted in a second
transition, presumably a $Q$-$K$ transition, and then approximated the curvature
of the crystal branch. The pressure remained higher than that of the perfect
crystal. The order parameters increase continuously after this freezing
transition. For $q\geq3.5$, we find that the $L$-$K$ transitions along the
growth and contraction branches approach one another. It is possible that
relaxation times are faster for these $q$ values and that transition to a liquid
state is more easily achieved.

\subsection{Phase diagram}

\label{sec:phase}

\begin{figure}
\includegraphics[width=0.6\textwidth, clip=true]{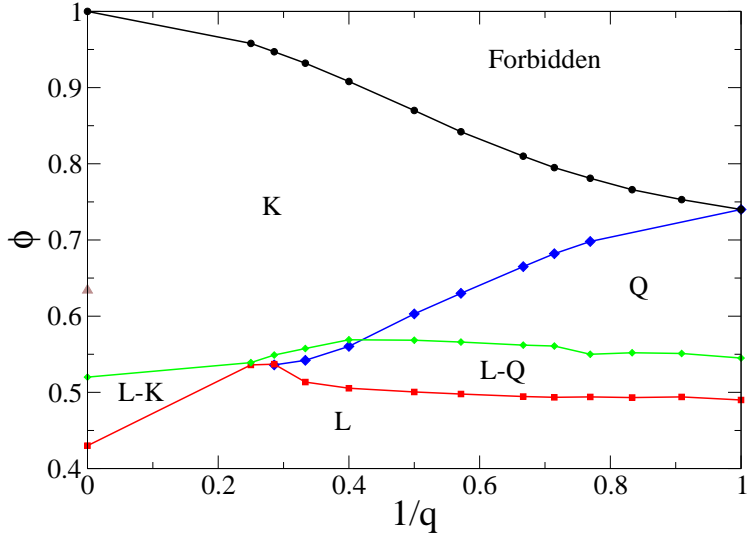}
\caption{(Color online) Approximate phase diagram for superballs spanning from
cubes to spheres. }
\label{fig:phasediagram}
\end{figure}

As with the case of spheres, we consider the density region between the $L$-$Q$
transition of the growth branch and the $Q$-$L$ transition of the contraction
branch to be coexistence between these two phases. In Fig.\
\ref{fig:crys.press.freeze}, this coexistence region is presumed to be the
region between sharp phase transitions along the growth and contraction
branches. Using the results of the DTS algorithm and the quantification of
orientational and translational order, the approximate phase diagram is shown in
Fig.\ \ref{fig:phasediagram}. 

The black circles represent the densities of the $\mathbb{C}_0$ and
$\mathbb{C}_1$ lattices at maximum packing. The blue diamonds represent the
density of the $K$-$Q$ transition as found along the crystal branch.  The
precise boundary was determined using a linear regression of data points on
either side of the transition and calculating their intersection.  The red
squares represent the first-order phase transition associated with the melting
of the crystal. It is the last density at which the crystal was stable in our
simulations. The green diamonds represent the freezing transition by allowing
particles to grow slowly. The data points along the cube line ($1/q = 0$)
represents the transitions identified by the authors of Ref.\ \cite{john2008pbc}
with the brown triangle being the crystal-cubatic transition they identified.  

While the $K$-$Q$ transition is monotonic in $q$, the $L$-$Q$ transition is not.
For $q\geq2.5$, the $K$-$Q$ transition of the contraction branch lies at a
density below the $L$-$K$ transition on the growth branch as in Fig.\
\ref{fig:p2.5.freeze}. The apparent tie line grows slightly as $q$ increases
from unity and begins to narrow significantly around $q=2.5$. By $q=4.0$, the
$K$-$L$ transition approaches the same density as the $L$-$K$ transition. There
are several possible explanations to the narrowing of the apparent coexistence
region near $q=4.0$.  We may not observe the possible coexistence region due to
the relaxation times in the system. The free-energy barrier associated with
melting may be reduced for increasing $q$ and thus the crystal may be difficult
to stabilize in a coexistence region.  Alternatively, we may be observing
microphase separation where the melting into a liquid phase is preferable to
maintain the crystal phase.  The system may also be large enough where certain
domains of the system are liquid while others remain crystal, although we do not
directly observe evidence for this phenomenon.  Further refinement of the phase
boundary would be necessary 
to answer these questions.

Detailed free-energy calculations are necessary to determine the precise phase
boundaries, refine the approximation of the phase diagram shown in Fig.\
\ref{fig:phasediagram}, and determine the coexistence pressures. One could
approximate the coexistence pressure using an equal area construction of the two
pressure branches. We expect that the coexistence pressures would be
nonmonotonic in $q$ as well.  The coexistence pressure for the sphere is
$Z=11.48$, and for the cube, $Z\sim12.5$. Using Fig.\ \ref{fig:p2.5.freeze} as
an example after rescaling the axes appropriately, one can estimate the
coexistence pressure to be much higher than 12.5. Ultimately, the free-energy
calculations would definitively determine such a conjecture.

\section{Structural Characteristics}
\label{sec:structure}

\subsection{Liquid Phase}
\begin{figure}
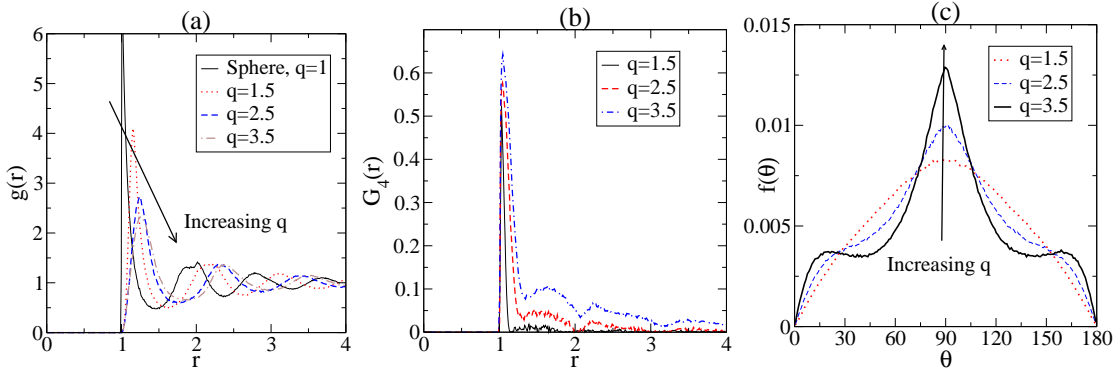

\includegraphics[width=0.28\textwidth, clip=true]{Fig10a.eps}
\includegraphics[width=0.3\textwidth,  clip=true]{Fig10b.eps}
\includegraphics[width=0.3\textwidth,  clip=true]{Fig10c.eps}
\caption{(Color online) (a) Radial distribution function $g_2(r)$, (b)
orientation correlation function $G_4(r)$, and (c) angular distribution function
$f(\theta)$ for various $q$ in the liquid phase at $\phi =0.54$ for $N=1000$.
The peak of the radial distribution function becomes smaller with increasing $q$
while the first neighbor peak moves farther due to excluded volume effects. The
local alignment increases with increasing $q$, but diminishes within four
particle diameters. Mutual alignment of neighboring particles increases sharply
for increasing $q$.} 
\label{fig:liq.rdf}
\end{figure}

The liquid phase of superballs lacks long-range orientational and translational
order but can show significant local order. Comparing systems at the same $\phi$
but for various values of $q$, we find that increasing $q$ yields increased
local orientational order but diminished local translational order.  The radial
distribution function for dense liquids at various $q$ for the common $\phi =
0.54$ is shown in Fig.\ \ref{fig:liq.rdf}(a). The first peak in the $g_2(r)$ is
reduced with increasing $q$, likely due to the interaction anisotropy building
up in the liquid. Because of the anisotropy of the particle, contacting
neighbors are not restricted to $r=1$ as they are for spheres, and thus the
first peak in $g_2(r)$ is lower than that for spheres.  
 
The strong peak in $G_4(r)$, defined by Eq.\ \ref{eq:ocf} at $r=1$ for all
particle shapes, shown by $G_4(r)$ in Fig.\ \ref{fig:liq.rdf}(b), shows that
particles with cube-like shape have a strong preference to align orthogonally
with contacting particles.  Evidently, the existence of cube-like shape is
sufficient to produce preferential cubatic alignment at contact, even for
particles with $q$ just above unity and whose cube-like shape is difficult to
see visually.  The degree of curvature determines the range of order in the
liquid phase.  For larger $q$, the orientational correlations persist for up to
three diameters.  Sharper corners effectively ``brace'' particles against
rotations.  The angular distribution function $f(\theta)$, Fig.\
\ref{fig:liq.rdf}(c), reinforces the notion of bracing. With sharper edges,
nearest neighbors have a greater preference to align along mutually orthogonal
directions in the liquid phase.
 
\subsection{High-Density Phases}

\begin{figure}
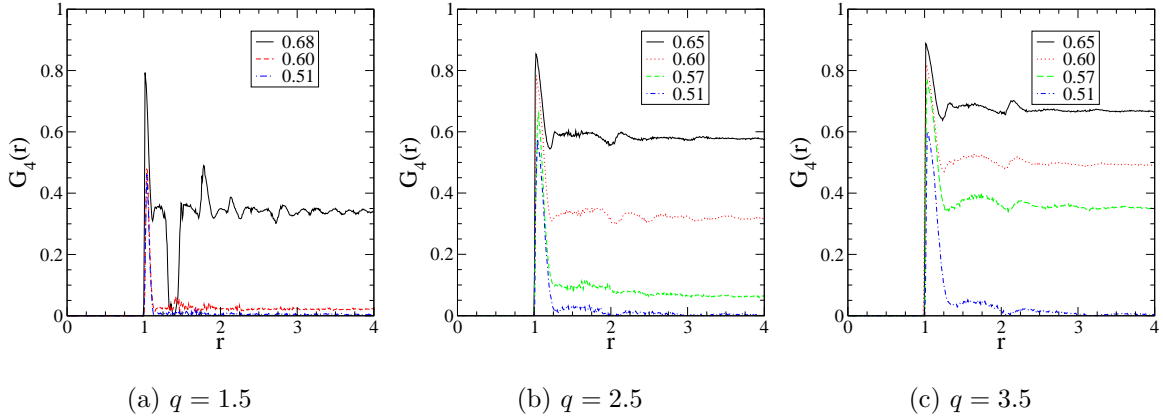

\subfigure[ $\text{ }q=1.5$]{\label{fig:g4.1.5}
\includegraphics[width=0.3\textwidth, clip=true]{Fig11a.eps}}
\subfigure[ $\text{ }q=2.5$]{\label{fig:g4.2.5}
\includegraphics[width=0.3\textwidth, clip=true]{Fig11b.eps}}
\subfigure[ $\text{ }q=3.5$]{\label{fig:g4.3.5}
\includegraphics[width=0.3\textwidth, clip=true]{Fig11c.eps}}
\caption{(Color online) Orientational correlation function $G_4(r)$ for various
$\phi$. } 
\label{fig:ocf.crys}
\end{figure}

While $S_4$ measures the global orientational order in the system, the
orientational correlation function $G_4(r)$ provides insight in the locality of
cubatic ordering. Figure \ref{fig:ocf.crys} shows $G_4(r)$ for several $q$
values. For $q=1.5$ in the $K$ phase, Fig.\ \ref{fig:g4.1.5} and $\phi = 0.68$,
the long-ranged orientational order is evident by the large value of $G_4(r)$
for nearly all $r$.  Particles with ``face-to-face'' contacts at $r=1$ and those
at pair distances associated with lattice sites are more strongly aligned than
those that deviate from lattice sites. However, those pairs of particles that
are separated by 1.35 diameters are misaligned, evidenced by $G_4(r)$ which is
nearly zero at this distance. Once the system enters the $Q$ phase,
orientational correlations remain local, although $S_4$ suggests there remains
some global cubatic order. These local orientational correlations are slightly
longer ranged than in the liquid phase, which results in a larger value of
$S_4$.   

For larger $q$, we observe similar trends, though particles are better
stabilized at all pair distances. The bracing of particles with sharper edges
prevent rotations at the pair distance of 1.35 diameters.  For $q=2.5$, Fig.\
\ref{fig:g4.2.5} shows that $G_4(r)$ has an initial maximum and minimum,
followed by weak maxima and minima associated with neighbor distances. The
general shape of the $G_4$ curve is maintained through all densities associated
with the $K$, aside from general loss of long-ranged order. In going from the
$K$ to $Q$ phase, the orientational correlations at long-ranged nearly vanish in
the same manner as that of $q=1.5$.  Fig.\ \ref{fig:g4.3.5} shows that the
orientational correlations for $q=3.5$ are similar to those at $q=2.5$. 
Surprisingly, the contact value of $G_4(r)$ was relatively unaffected by the
density in all cases. The degree of curvature and density have little influence
on the ordering at contact, but rather it is the presence of cubic symmetry that
produces alignment at $r=1$.  

\begin{figure}
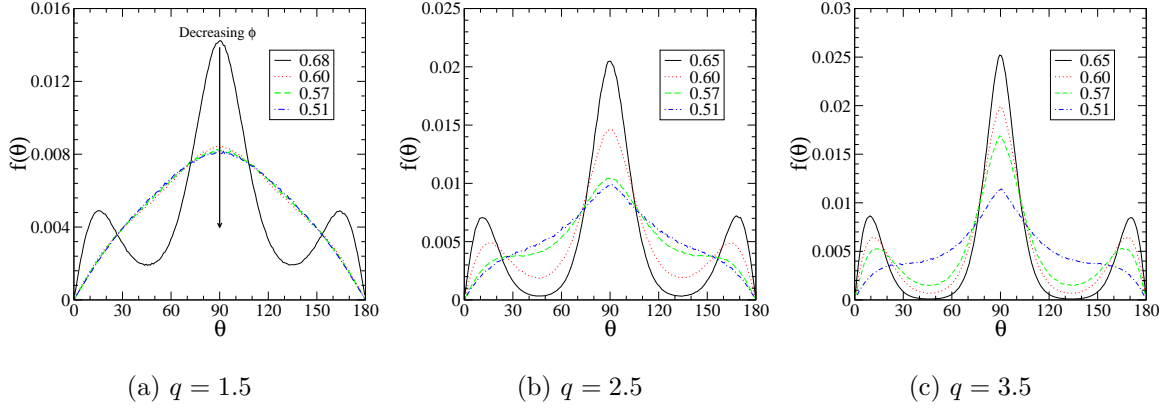

\subfigure[ $\text{ }q=1.5$]{\label{fig:ang.1.5}
\includegraphics[width=0.3\textwidth, clip=true]{Fig12a.eps}}
\subfigure[ $\text{ }q=2.5$]{\label{fig:ang.2.5}
\includegraphics[width=0.3\textwidth, clip=true]{Fig12b.eps}}
\subfigure[ $\text{ }q=3.5$]{\label{fig:ang.3.5}
\includegraphics[width=0.3\textwidth, clip=true]{Fig12c.eps}}
\caption{(Color online) Angular distribution function $f(\theta)$ for various
$\phi$.}
\label{fig:adf.crys}
\end{figure}

To further examine the local ordering environment, we used the angular
distribution function $f(\theta)$, shown in Fig.\ \ref{fig:adf.crys} for several
$q$. This measures the correlations between the alignment of principal axes for
neighboring particles. Local cubatic ordering is present when the function has
three maxima but has minimal cubatic ordering when there is a single maximum.
When initialized, the crystals are perfectly aligned and the angular
distribution function is $f(0^\circ)=1/3$ and $f(90^\circ)=2/3$. Since the axes
are labelled ({\it i.e}.\ all of the $A$ axes are in the same direction), the
crystal system will have an $f(\theta)$ that is asymmetric in $\theta$ at high
densities. At equilibrium, particles will have flipped sufficiently to mix the
labels and $f(\theta)$ will become symmetric about $90^\circ$.  The data shown
in Fig.\ \ref{fig:adf.crys} illustrates that this symmetry is easily achieved at
these densities even in the $K$ phase. 

For $q=1.5$, Fig.\ \ref{fig:ang.1.5} shows that superballs in the $K$ phase
maintain strong neighbor bracing. Orthogonal alignment among neighbors is the
most probable alignment. Entering the $Q$ phase reduces the peak at
$\theta=90^\circ$ and eliminates the other local maxima.  The shape of the curve
is similar to that of the dense liquid phase.  

For $q=2.5$, Fig.\ \ref{fig:ang.2.5} demonstrates that the $K$ phase prevents
nearly all neighboring particle axes from aligning at $45^\circ$ since the depth
of the local minima approach zero.  As the $K$-$Q$ transition is neared,
$\phi=0.57$, the shape of the curve changes distinctly as the local minima and
maxima converge.  For further increases in $q$, Fig.\ \ref{fig:ang.3.5} shows
that the peaks become sharper and narrower, though the fundamental shape does
not change.   

\begin{figure}
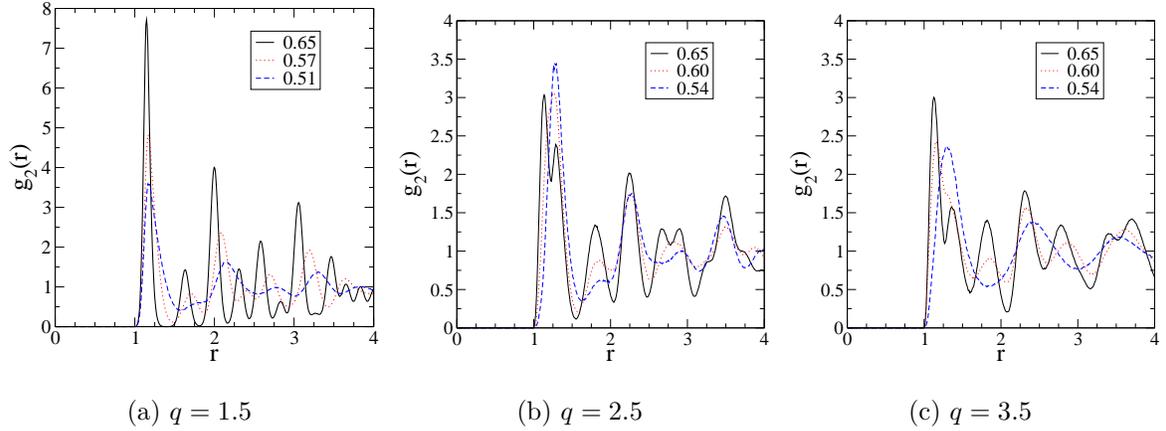

\subfigure[ $\text{ }q=1.5$]{\label{fig:gr.1.5}
\includegraphics[width=0.3\textwidth, clip=true]{Fig13a.eps}}
\subfigure[ $\text{ }q=2.5$]{\label{fig:gr.2.5}
\includegraphics[width=0.3\textwidth, clip=true]{Fig13b.eps}}
\subfigure[ $\text{ }q=3.5$]{\label{fig:gr.3.5}
\includegraphics[width=0.3\textwidth, clip=true]{Fig13c.eps}}
\caption{(Color online) Radial distribution function for various $\phi$. } 
\label{fig:rdf.crys}
\end{figure}

The radial distribution function $g_2(r)$ generally shows that translational
order is maintained thorough the $K$ and $Q$ phases.  The crystal-like
characteristics, peaks in $g_2(r)$, are evident but smear out gradually as the
density is reduced.  Fig.\ \ref{fig:rdf.crys} shows $g_2(r)$ for various $q$ and
several densities.  While the peaks in the liquid phase, Fig.\ \ref{fig:gr.3.5}
shows that for $\phi=0.54$, decay quickly to unity, the peaks in the $K$ and $Q$
phases do not appear to decay rapidly. Evident in $g_2(r)$ for $q=2.5$ and 3.5
is a split first peak that represents the subtle difference in pair distances of
the first and second neighbors associated with the corresponding $\mathbb{C}_1$
lattices.  In addition, the particle distribution functions, in which the
particle coordinates are projected onto a line perpendicular to the crystal
planes, are completely periodic throughout the $K$ and $Q$ phases.  This suggest
that lower-dimensional translational order, such as that in columnar phases, was
not present in either the $K$ or $Q$ phases.
 
\section{Discussion}
\label{sec:discuss}
 
In this paper, we have determined the phase behavior of a general class of hard
convex particles with shapes between a sphere and cube.  We found that the
degree of curvature plays a significant role in the cubatic ordering and phase
transitions. Despite seemingly similar visual appearance among particle shapes,
subtle differences in the curvature can yield large changes in the EOS.  In the
liquid phase, edges and corners are the dominant features that contribute to
excluded volumes, while in the crystal phase, they can brace against rotations.

Although MD has some advantages over Monte Carlo techniques, including the
ability to simulate cooperative behavior more efficiently, there are always
questions concerning kinetic effects and boundary effects. We have performed
comprehensive tests to address and minimize these issues. The use of a
deformable box helped to alleviate the possibility of anisotropic stresses and
to reduce boundary effects. In each case, we verified that the pressure tensor
remained isotropic throughout the simulations.  The pressure tensor showed
slight anisotropy immediately before the melting transition, likely due to the
large stresses that build up and induce the transition. In all other cases,
including $K$-$Q$ transitions, the pressure tensor remained isotropic. We tested
several system sizes within the limits of our computational capabilities and we
found that the curvature of the EOS remains unchanged as the system size was
increased. Smaller systems produced larger fluctuations and slight variations in
the density of the melting transition compared with larger systems. 

In addition, we tested systems in what we knew to be poor initial conditions.
For example, for $q=2.0$, we initialized the system in three distinct crystals -
face-centered cubic (FCC), simple cubic (SC), and the $\mathbb{C}_1$ crystal.
The FCC and SC cases nearly converged to the pressure equation of state
associated with the $\mathbb{C}_1$ crystal.  This supports the notion that we
have used simulation conditions that can approximate the equilibrium phase
behavior. The internal stresses of the FCC and SC simulations were relieved when
the system shifted toward the $\mathbb{C}_1$ crystal.

As with many simulation methodologies, kinetic effects are possible due the
inability to simulate long-time behavior.  We tested several growth rates and
found little variation among the EOS's generated for rates of $|\gamma|<
10^{-5}$. Long, constant-density trajectories revealed slight variation in the
densities associated with freezing and melting transitions, though not
substantial enough to warrant concern. The curvature of the EOS in the $K$ and
$Q$ phases were insensitive to the growth rates $|\gamma|< 10^{-5}$.

The primary extension of this work includes the refinement of the boundaries in
the phase diagram. As we have shown in Fig.\ \ref{fig:phasediagram}, the DTS
algorithm has shown first-order phase transitions quite well. However, detailed
free energy calculations are needed in order to provide precise phase boundaries
and identify higher-order phase transitions, if they exist. In addition, we have
yet to explore the dynamic behavior in hard superball systems.  The diffusion
coefficient may exhibit discontinuous jumps when crossing between $K$ and $Q$
phases. Additionally, rotational degrees of freedom play an important role in
the glassy phase. Understanding the extent of curvature of corners and edges on
the jamming characteristics has undergone initial exploration
\cite{jiao2010dfa}. It may also be possible to manufacture colloidal particles
with such controlled shape via photolithography or other synthetic techniques. 
Testing these systems for certain technologically relevant properties including
wave scattering characteristics and rheology may reveal unusual behavior. 

While we have used a particle-growth algorithm to understand the phase behavior
of hard particles,  particle-growth algorithms are often used to search for
optimal particle packings. In our experience, allowing particles in a dense
liquid to grow slowly generally produces a partially crystalline system. Our
results identify one of the primary challenges associated with searching for
optimal packing arrangements since these algorithms could hardly ever achieve
the densest state. The relaxation times are far too long for computer
simulations. This highlights the need for alternative methods to find dense
packings of particles \cite{jiao2008ops, jiao2009ops, torquato2009dense}. Only
after finding the densest packings can researchers attempt to determine the
entire equilibrium phase behavior of hard particles by applying particle
contraction.

The application of overlap potentials to generalized convex particles
\cite{donev2006thesis} and the development of efficient MD algorithms
\cite{donev2005neighbor1, donev2005neighbor2} has made available the opportunity
to explore more deeply how shape influences phase behavior.  Along these lines,
planned future work includes determining the onset of nematic, smectic, and
possibly parquet phases of elongated superballs. This particular perturbation
would allow one to explore the continuous evolution from ellipsoids to
tetragonal parallelepipeds, which exhibit various types of liquid crystals for
certain aspect ratios.  With the tools available, one can determine, for
example, where the crossover point is for the appearance of a parquet phase in a
system of elongated superballs. Additionally, a study of parallel hard
superellipsoids, a perturbation from spheres or ellipsoids to cylinders, showed
the onset of a smectic phase \cite{martinez2008nlc}. With the addition of
rotational degrees of freedom, this system that would presumably exhibit a
cubatic or parquet phase depending on the deformation parameter.  There are
seemingly endless possibilities for hard-particle shapes and perhaps a
mathematical treatment of generalized particles with ``superexponents'' may be
enlightening.

\section*{Acknowledgments}
The authors are indebted to the reviewer of an earlier manuscript that
recommended verifying the isotropy of the pressure tensor. The authors thank
Aleksandar Donev for making his molecular dynamics codes publicly available and
helpful discussions.  The authors gratefully acknowledge Fernando Escobedo for
providing the hard cube EOS data.  R.D.B.\ thanks Yang Jiao for valuable
discussions. S.T.\ thanks the Institute for Advanced Study for its hospitality
during his stay there. This work was supported by the Office of Basic Energy
Sciences, US Department of Energy,
Grant No. DE-FG02-04-ER46108.

\appendix

\section{Phase Behavior Effects Induced by System Geometry}
\label{sec:appendix}

Here we report on the importance of utilizing a deformable boundary in
simulations of anisotropic particles.  Because superballs are centrally
symmetric objects, we expected that the pressure tensor would remain isotropic
throughout simulations. The EOS for the liquid does not depend on the geometry
of simulation box. The pressure tensor for the liquid is isotropic since
particles rotations are random. 

\begin{figure}
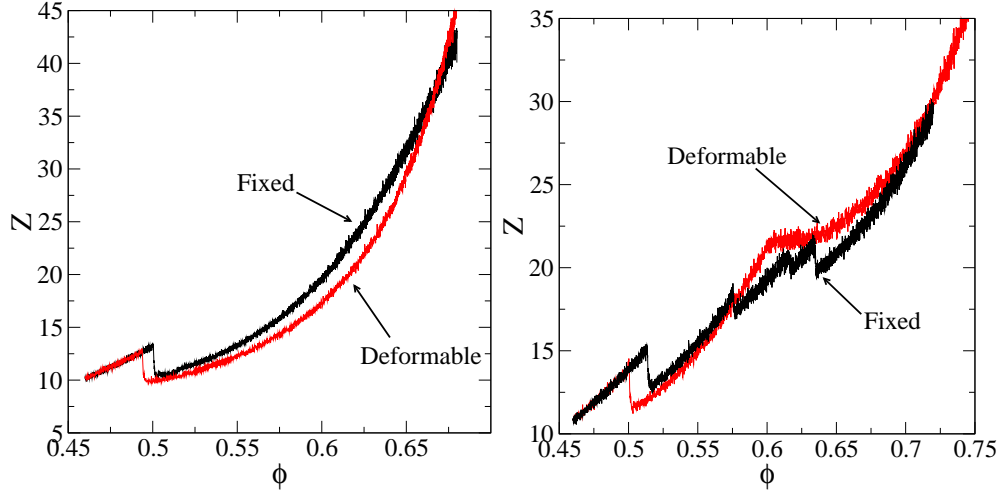

\includegraphics[width=0.39\textwidth, clip=true]{Fig14a.eps}
\includegraphics[width=0.4\textwidth,  clip=true]{Fig14b.eps}
\caption{(Color online) Comparison of equation of states obtained using fixed
and deformable system boundaries for (left) $q=1.3$ and (right) $q=2.0$. For
small $q$, the curvature is sharper with a deforming boundary than with a fixed
boundary. For larger $q$, a series of apparent phase transitions are induced due
to the ability of particles to rotate about certain axes.} 
\label{fig:compare}
\end{figure}

However, in high-density phases, we find that anisotropic stresses can build up
when the boundaries are fixed, even in the case where $q$ is near unity.  Fig.\
\ref{fig:compare} compares the equations of state generated using a fixed box
and a deforming box for $q = 1.3$ and 2.0.  For small $q$, the equations of
state differ significantly in their curvature. 

For $q\geq1.4$, a fixed simulation box not only affects the curvature but also
introduces a series of apparent first-order phase transitions.  As seen in Fig.\
\ref{fig:compare} for $q=2.0$, the system with a fixed simulation box has three
apparent first order phase transitions. These discontinuities along the branch
are highly reproducible. Each apparent phase transition can be attributed to
rotations of particles about certain axes.  The highest-density transition
corresponds to rotations about a single particle axis. At this density, there is
enough average spacing between particles in the lattice to allow for rotations
about a single axis.  The middle transition corresponds to rotations about two
particle axes while the lowest-density transition on the crystal branch
corresponds to free rotations about three particle axes.  In general, the
melting densities in the deforming box and fixed box differed slightly, with the
deforming box having a lower melting density.  In all cases, melting appeared to
be a first-order phase transition. 

Observing the elements of the pressure tensor as a function of density shows
that anisotropic stresses build up as the system approaches each apparent phase
transtion. With deforming box simulations using the Parrinello-Rahman-like
algorithm allows for these internal stress to be relaxed away quickly, as they
would occur in thermodynamic systems.  Although the Parrinello-Rahman-like
algorithm does not rigorously sample an isostress ensemble because the lattice
vectors are not directly coupled to the particle interactions, it is a
reasonable approximation.

\newpage


\begin{thebibliography}{57}
\bibitem[{foo({\natexlab{a}})}]{footnote2}
\bibinfo{note}{We use {\it q} for the deformation parameter whereas previous
  literature has used {\it p} since we reserve that character to represent the
  pressure.}
\bibitem[{foo({\natexlab{b}})}]{footnote}
\bibinfo{note}{Equation of state data were provided by Fernando Escobedo and
  were generated by Monte Carlo simulation of 1728 cubes in the
  isobaric-isothermal ensemble.}
\end{thebibliography}
\end{document}